\documentclass[review,number,amssymb,amsthm,floatfix]{elsart}
\usepackage{lineno}
\input epsf
\begin{document}

\begin{frontmatter}
\title{Drift Chamber Alignment using Cosmic Rays}
\author{Ashutosh V. Kotwal\thanksref{cor}} 
\thanks[cor]{Corresponding author. Tel: (919) 660-2563; fax: (919) 660-2525; E-mail address: ashutosh.kotwal@duke.edu}
\address{{\em Duke University} }
\author{Christopher P. Hays}
\address{{\em Oxford University} }
\begin{keyword}
drift chamber \sep alignment \sep cosmic ray \sep calibration.
\PACS 06.60.Sx \sep 06.30.Bp \sep 07.05.Kf \sep 07.77.Ka  \sep 07.90.+c 
\sep 29.30.Aj \sep 29.40.-n \sep 29.40.Gx \sep 29.40.Cs \sep 29.90.+r 
\end{keyword}

\begin{abstract}
The Collider Detector at Fermilab (CDF) is a general-purpose experimental apparatus 
with an inner tracking detector for measuring charged particles, surrounded by a 
calorimeter for measurements of electromagnetic and hadronic showers, and a muon detector
 system.  We present a technique for, and results of, a precise relative alignment of the drift chamber
 wires of the CDF tracker. This alignment has been an important component of the track momentum calibration, 
 which is the basis for the charged-lepton calibration for   
the measurement of the $W$ boson mass at CDF. 
\end{abstract}

\end{frontmatter}

\section{Introduction}
\label{sec:intro}
The measurement of the $W$ boson mass with the CDF detector~\cite{wmassPRD} at the Fermilab Tevatron 
 $p \bar{p}$ collider  
 achieves a precision of 
 one part per 10,000 on the measured energy of muons from $W$ boson decays~\cite{cdfwmass}.  
A key component of the momentum  calibration is a precise relative alignment of the wires in the CDF drift chamber. In this paper we describe the alignment technique developed
 at CDF using cosmic rays collected {\it in situ} with collider operation. This technique was initiated in 2002 and used to perform the alignment for the first two
 measurements of the $W$ boson mass from the CDF Run II data, based on 200 pb$^{-1}$~\cite{wmassPRD} and 2.2 fb$^{-1}$~\cite{cdfwmass} of integrated luminosity, respectively.
  We also present the results from the alignment analysis performed most recently using cosmic rays collected during
 the entire Run II collider operation, to be used in the next $W$-boson mass measurement at CDF. 

The momentum calibration~\cite{cdfwmass} is performed in two steps.  First, the selected cosmic-ray sample is reconstructed using special pattern-recognition and 
 track-fitting algorithms  
 that reconstruct the complete cosmic-ray trajectory through both halves of the drift chamber. 
 These algorithms and their performance are described in~\cite{cosmicNIM}.   
 This fitted trajectory proves an excellent reference, with respect to which the average hit 
 residuals can be minimized to achieve an internal alignment of the wires. Certain degrees of freedom associated with deformations of the tracker endplates cannot be constrained
 by minimizing residuals with respect to collider tracks. 
 Examples of such deformations are relative rotations of the inner and outer cylinders of the 
 drift chamber, and relative twists of the east and west endplates. They result in parameter biases in collider track fits but do not result in displaced residuals. On the other hand, 
 such deformations result in differences between the trajectories on the two sides of the drift chamber of the same cosmic ray, and can be constrained by using the complete two-sided helix fit. 
 This property is exploited to remove a number of important
 sources of bias in the measurement of collider tracks. 

 In the second step of the calibration, $J/\psi \to \mu \mu$ and $\Upsilon \to \mu \mu$ mass peaks are reconstructed with the above alignment. Using the precisely known masses of these particles, 
 a momentum scale factor and the ionization energy loss model are tuned simultaneously  in the simulation,
  and then applied to the $W$ boson mass measurement~\cite{wmassPRD}.  Residual misalignments that cannot be constrained with cosmic-ray
 tracks are corrected using the ratio of calorimeter energy to track momentum of electrons and
 positrons from $W$-boson decays~\cite{wmassPRD,cdfwmass}.

In the following we describe the procedure and the results obtained from the cosmic ray alignment. 
In Sec.~\ref{sec:COT} we provide a summary of the CDF drift chamber construction and the degrees of freedom we allow in the alignment procedure. 
 In Sec.~\ref{sec:Distributions} we describe the spatial and kinematic distributions of the 
 cosmic ray sample recorded with the CDF detector, and the sample selection. In Sec.~\ref{sec:Corrections} we 
 show comparisons of the position residuals before and after the alignment procedure is performed. 
 Corrections to the wire shape due to gravitational sag and electrostatic deflection are discussed in Sec.~\ref{sec:wireShape}. 
 The tuning of the drift model is discussed in Sec.~\ref{sec:driftModel}. 
 Finally, in Sec.~\ref{sec:Biases}
  we show the reduction in the track parameter biases, as determined from the comparison 
 of the two segments of the cosmic ray track. 
\section{Alignment degrees of freedom}
\label{sec:COT}
\begin{figure*}[!htp]
\begin{center}
\epsfxsize 13.0cm
\epsffile {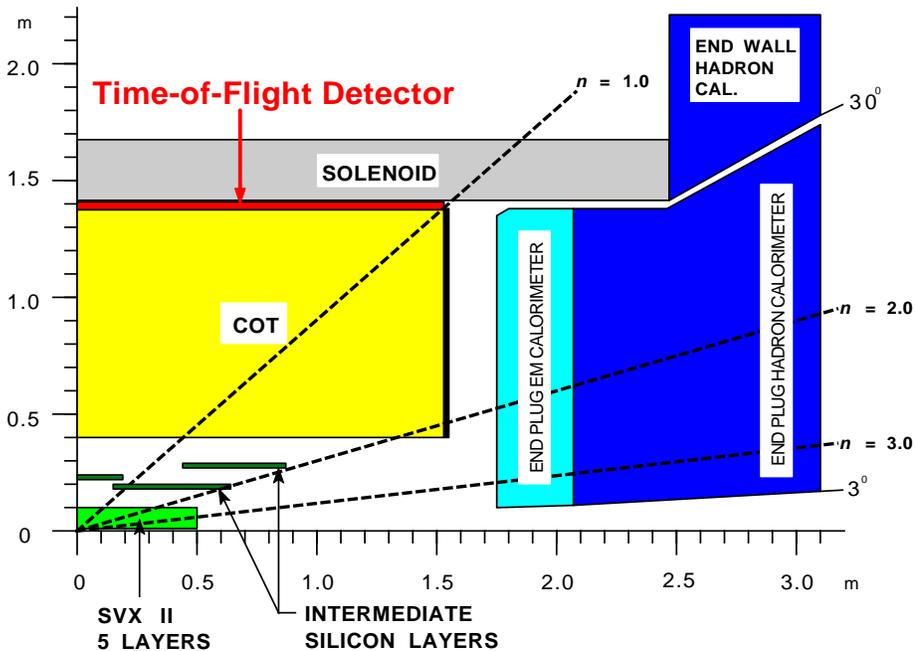}
\caption{A cut-away view of the CDF detector, reproduced from~\cite{cot}. Not shown are the barrel calorimeters outside the solenoid, and the muon detectors outside the calorimeters.  }
\label{fig:cdf}
\end{center}
\end{figure*}

The CDF detector~\cite{wmassPRD,cdfNIM,jpsiReference,xsecPRD}  is shown in Fig.~\ref{fig:cdf}. The   central tracking drift 
chamber~\cite{cot} (COT) uses an open-cell geometry with 30,240 sense wires. Its tracking volume extends from an inner radius of 41 cm to an outer 
 radius of 138 cm, with a 
 longitudinal extent of 310~cm. A superconducting solenoid immediately outside the  COT provides a nearly uniform 1.4~T magnetic field in the tracking volume.
 Within the COT sits a silicon detector to provide precise vertexing information.
 In the measurement of the $W$-boson mass, the momentum of the charged lepton produced promptly in the
 collision is measured using COT information and constrained to the beam collision region in the
 transverse plane; silicon detector hits do not improve the precision on the beam-constrained track parameters  and are not used. 

\begin{figure*}[!tbhp]
\begin{center}
\epsfxsize 12.0cm
\epsffile {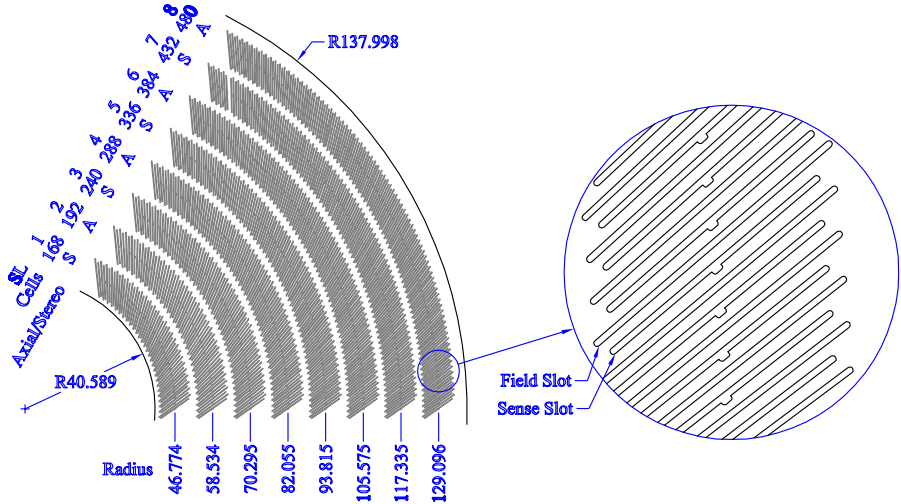}
\caption{A section of an aluminum endplate of the COT, reproduced from~\cite{cot}. The slots cut in the endplates anchor individual drift cells containing 12 sense wires each.  }
\label{fig:cot}
\end{center}
\end{figure*}

 A section of one of the aluminum endplates of the COT is shown in Fig.~\ref{fig:cot}. The drift chamber consists of 2520 drift cells, each containing 12 sense wires. The 
 maximum drift distance is $\approx 8.8$~mm. The cells are arranged in 8 radial superlayers ({\sc sl0} to {\sc sl7}), with the number of cells per superlayer increasing with radius. 
 Alternating superlayers consist of wires running along the longitudinal axis (axial superlayers) and wires with a $\pm 2^\circ$ stereo angle (stereo superlayers).  The stereo angle 
 changes sign from one stereo superlayer to the next. 

\begin{figure*}[!tbhp]
\begin{center}
\epsfxsize 10.0cm
\epsffile {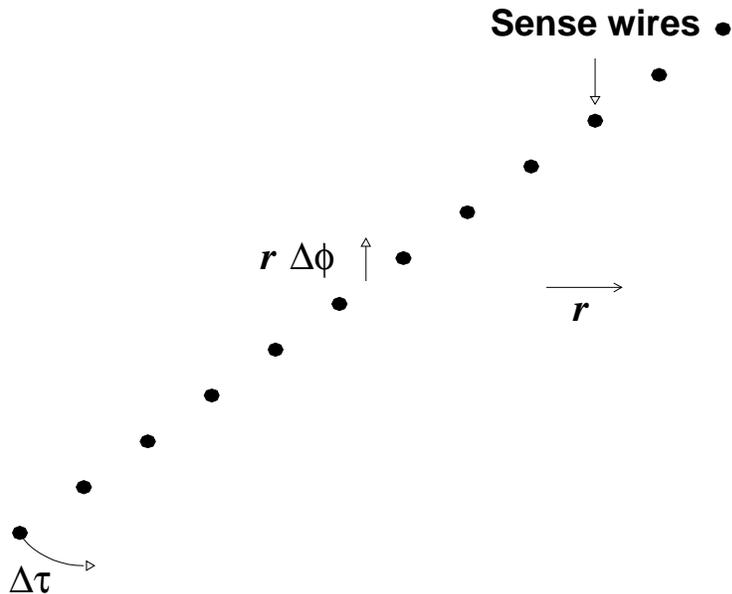}
\caption{A drift cell showing 12 sense wires, along with the radial ($r$) coordinate, and azimuthal ($r \Delta \phi$) and tilt ($\tau$) correction parameters.  }
\label{fig:cotcell}
\end{center}
\end{figure*}

The radial spacing between sense wires in a cell is 5.8~mm~\cite{cot}. The wires are attached at their ends to rigid cards which are precision-mounted on the COT endplates. In 
 the  alignment model, 
 each cell's profile at the endplates is described by a straight line (see Fig.~\ref{fig:cotcell}). Thus, the degrees of freedom to be constrained 
 in order to precisely locate  each sense wire at each endplate are the following:
\begin{enumerate}
\item the transverse $(x,y)$ coordinates of the center of each cell, at the longitudinal ($z$) coordinate $\pm 155$~cm of the two endplates. 
\item the tilt angle $(\tau)$ of each cell relative to the radial vector from the transverse origin to the center of the cell at $z = \pm 155$~cm.
\end{enumerate}
 We parameterize the former degrees of freedom in terms of  symmetrized (i.e. averaged over the two endplates) and anti-symmetrized (i.e. difference between the
 two endplates) cell-center coordinates. The advantage of these definitions is that the symmetrized and anti-symmetrized cell-coordinate residuals are, to a large extent,  uncorrelated because
 of the approximately uniform and symmetric distribution of the cosmic rays in the $z$-coordinate. The symmetrized cell-coordinate residuals are averaged over $z$,  while the anti-symmetrized
 corrections are sensitive to the dependence of the hit residuals on the $z$-coordinate. 

 The cell-tilt angle ($\tau$) is designed to account for the Lorentz angle 
 of the drift direction given the magnitude of the electric field, the spectrometer magnetic field and the drift speed. The alignment
 corrects for small deviations in the cell tilt. The corrections are found to be almost the same for the two endplates. We compute the average correction to the cell tilt since the difference
 between the tilt corrections for the two endplates has negligible impact on tracking biases. 

 Finally, the functional forms describing the wire shape as a function 
 of $z$, and its variation with azimuth and radius, are also tuned. These degrees of freedom are discussed in Sec.~\ref{sec:wireShape}. 
\section{Cosmic ray distributions and sample selection}
\label{sec:Distributions}
\begin{figure*}[!tbhp]
\begin{center}
\epsfxsize 17.0cm
\epsffile {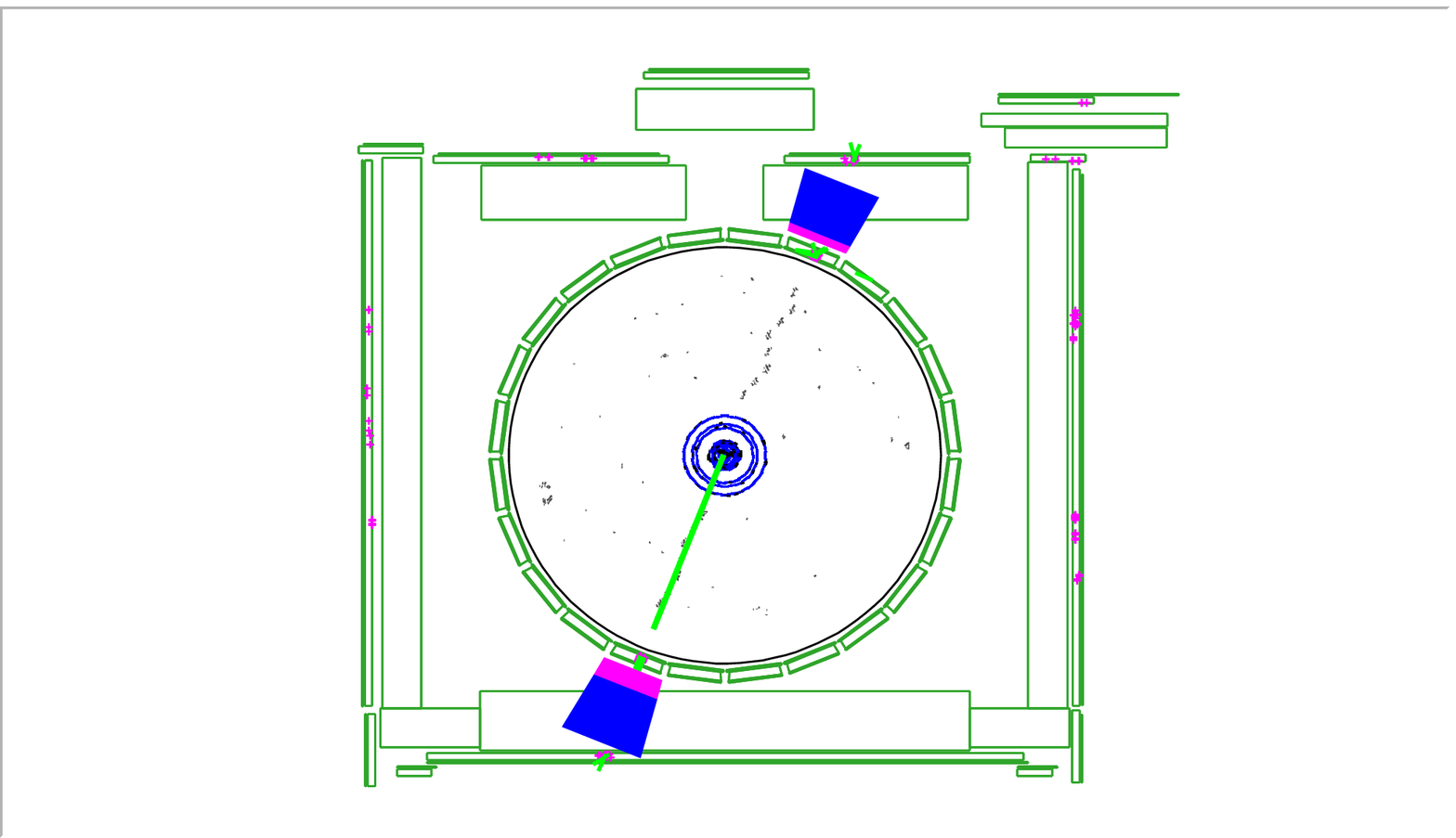}
\end{center}
\caption{Display of a cosmic-ray event recorded in coincidence with a beam crossing, in the absence of a $p \bar{p}$ collision. 
 The reconstructed helical track trajectory shown in the bottom half of the chamber
is found using the standard CDF tracking algorithm.  The top half of the trajectory is found using
the dedicated cosmic-ray reconstruction algorithm~\cite{cosmicNIM}, which also combines all the hits into a single dicosmic track. The reconstructed track has $p_T \approx 69$~GeV
 and $\eta \approx 0.2$. The COT hits are shown at $z=0$, resulting in  a staggering of displayed hits in stereo superlayers. }
\label{fig:cosmicRay}
\end{figure*}

The cosmic ray sample used in this study is collected during collider operation by
  high-$p_T$~\cite{conventions} muon physics triggers, ensuring that the cosmic rays and collider tracks are recorded under the same
 operating conditions for the drift chamber 
 and the spectrometer.  In order to isolate a clean and unambiguous sample of cosmic rays, we require that only one or two tracks be reconstructed
 in the event, and that at least one of them is tagged as a muon by the presence of a matching track segment reconstructed in the muon detectors. These criteria 
 efficiently select those cosmic-ray events 
  which coincide in time with beam crossings in which no proton-antiproton collisions occurred. A display of such a cosmic-ray event is shown in Fig.~\ref{fig:cosmicRay}. After all requirements,
 a sample of 207,023 cosmic-ray events is selected for the alignment study. 

\begin{figure*}[!tbhp]
\begin{center}
\epsfxsize 6.5cm
\epsffile {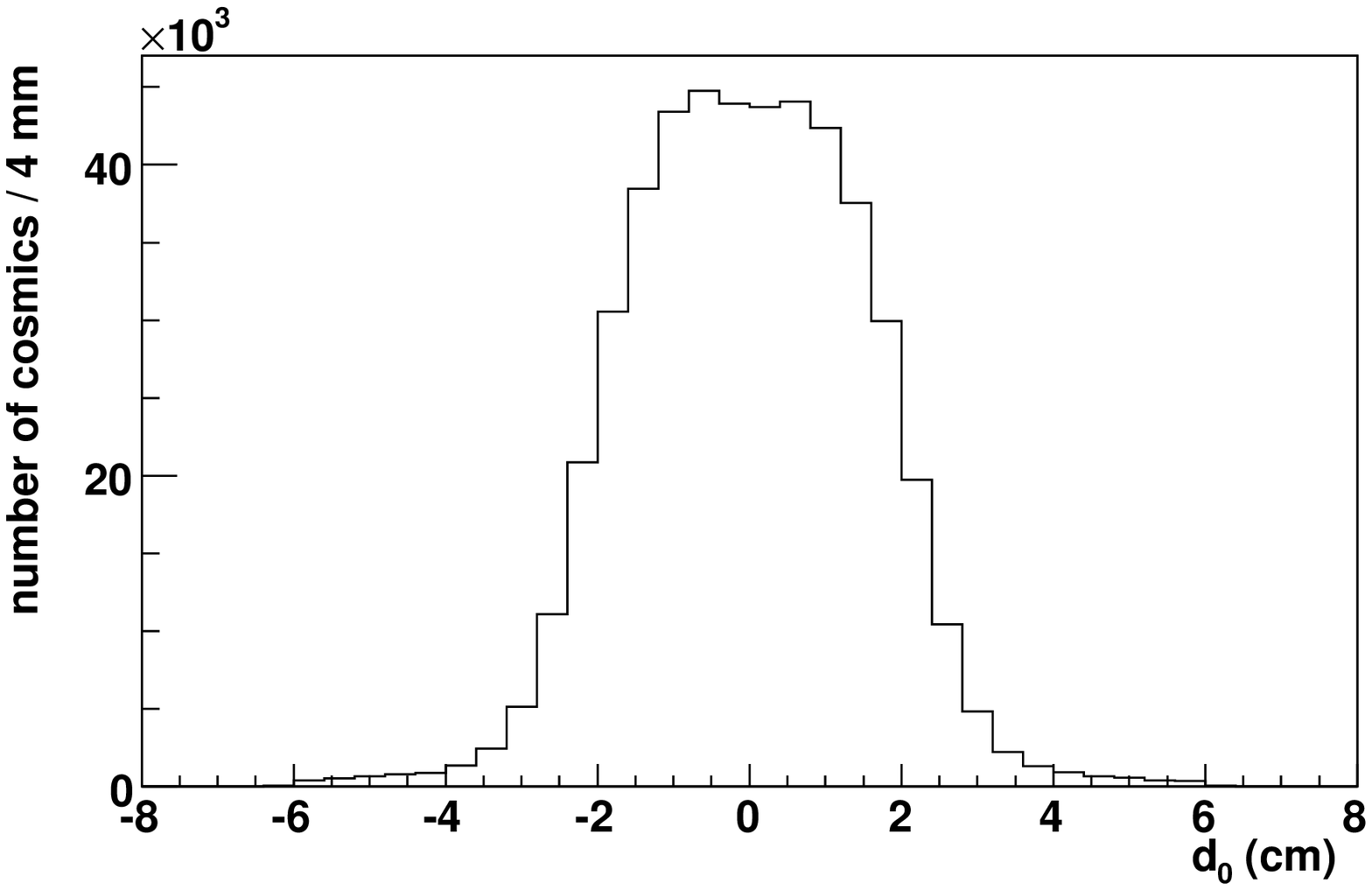}
\epsfxsize 6.5cm
\epsffile {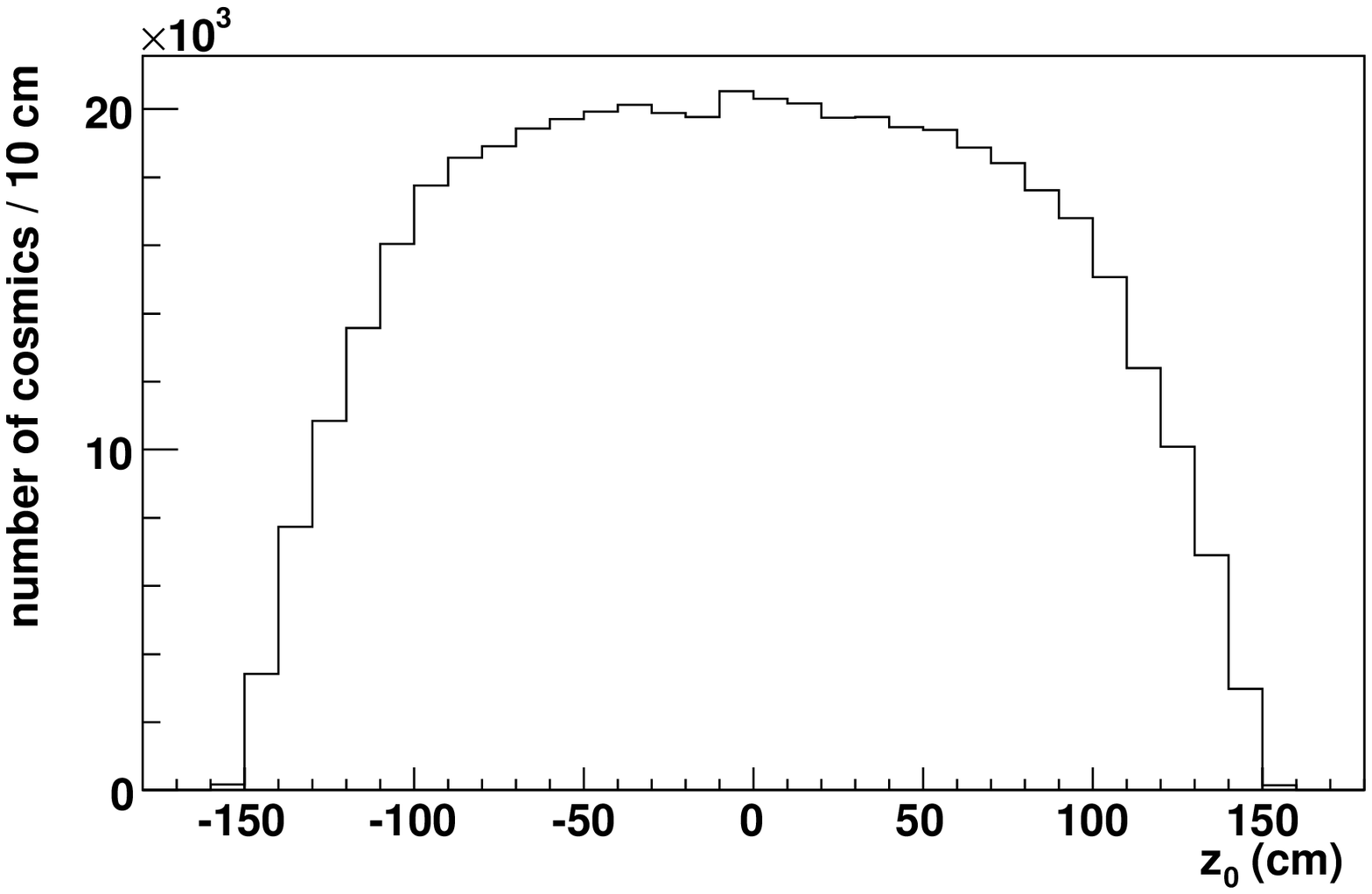}
\epsfxsize 6.5cm
\epsffile {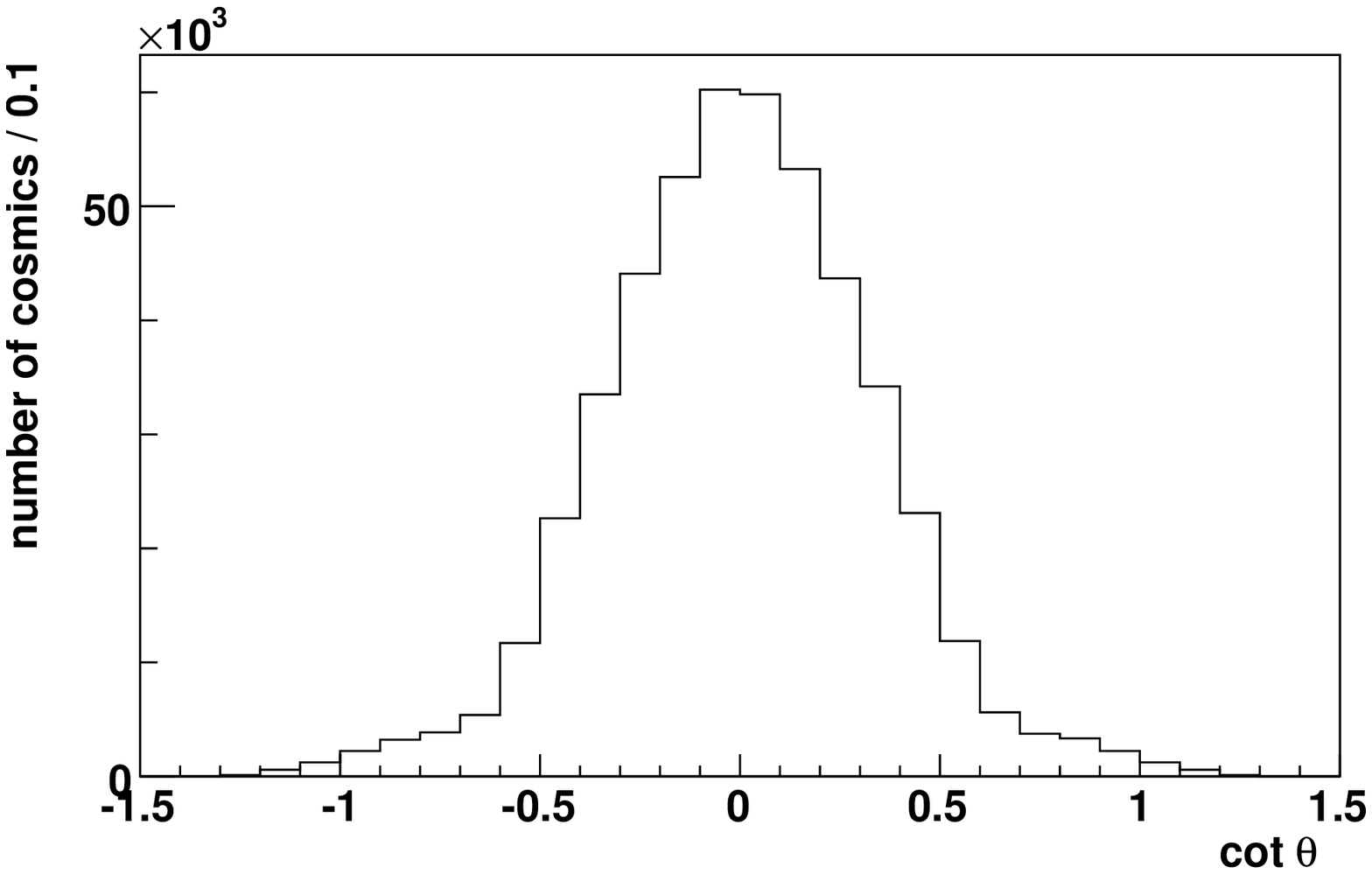}
\epsfxsize 6.5cm
\epsffile {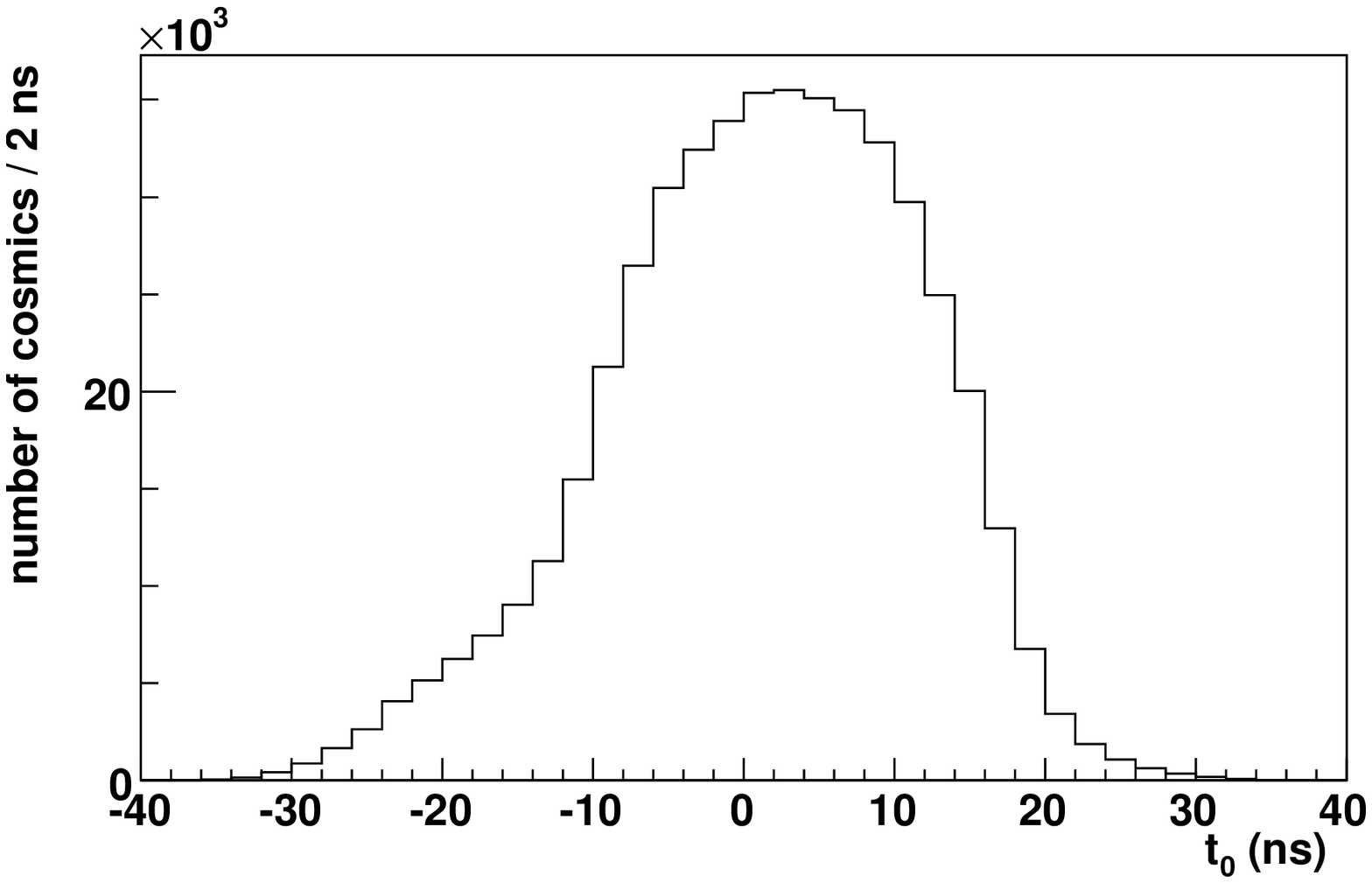}
\end{center}
\caption{Distributions of (top left) the transverse impact parameter $d_0$ with respect to the beam line; (top right) the longitudinal coordinate $z_0$ of the cosmic ray track 
 at the point of closest approach to the beamline; (bottom left) cotangent of the polar angle; and (bottom right) the time difference $t_0$ between the beam crossing time and the passage of the
 cosmic ray. With all criteria applied except for the $z_0$ and hit requirements, the sample contains 484,513 cosmic-ray events.   }
\label{fig:Distributions}
\end{figure*}

\begin{figure*}[!tbhp]
\begin{center}
\epsfxsize 6.5cm
\epsffile {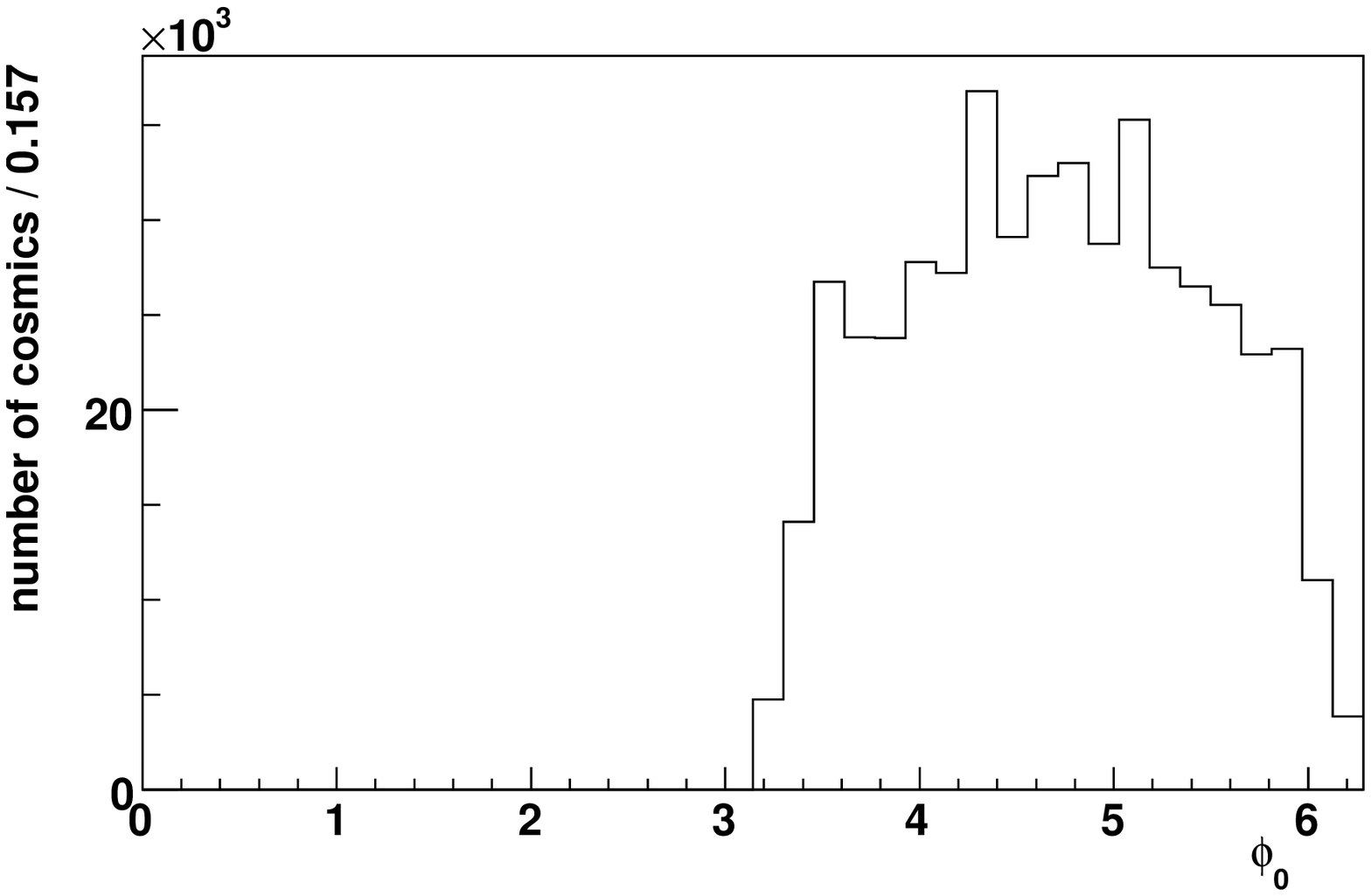}
\epsfxsize 6.5cm
\epsffile {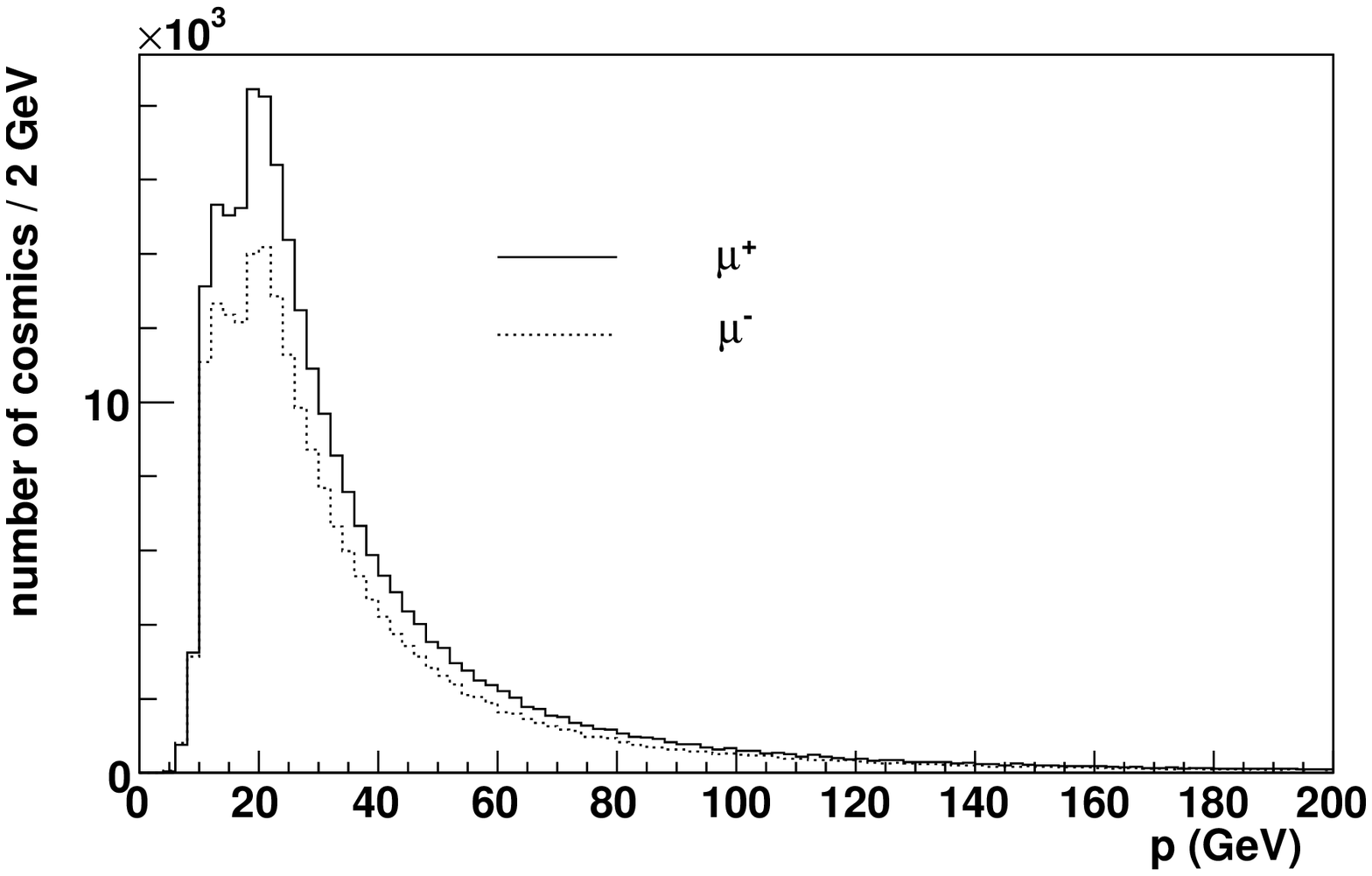}
\end{center}
\caption{Distributions of (left) the azimuthal direction of propagation $\phi_0$ and (right) the momentum of the cosmic ray tracks.   }
\label{fig:phiMom}
\end{figure*}

The spatial and temporal distributions of the cosmic rays are shown in Fig.~\ref{fig:Distributions}. The distributions of the azimuthal direction of propagation and the momentum are
 shown in Fig.~\ref{fig:phiMom}.

 The sample contains about 20\% more positively charged than negatively
charged muons, with similar momentum distributions between the two.  The residuals with respect to the
 two-sided helical track fit (referred to as the ``dicosmic track'') are an unbiased measurement of relative misalignments of the sense wires if and only if the true 
 parameters of the cosmic
 ray trajectory are the same on the two sides of the COT. In this case, fitting the hits on both sides with a single helix represents a valid model of 
 the muon's trajectory. In practice, 
 the muon loses on average $\approx 20$~MeV of energy while 
  passing through the silicon tracking detector. As a result, the latter half of the trajectory
 has lower momentum and the hit residuals will be biased with respect to the dicosmic track. Fortunately, this bias is in opposite directions for positive and 
 negative tracks. We weight the
 positive and negative muons in inverse proportion to their relative rates such that  the average curvature (defined as $q / p_T$ where $q$ is the charge and $p_T$ is
  the transverse momentum)
 of the reweighted combined sample is zero. As a result, the average residuals correspond to those of infinite-momentum tracks, and are not influenced by the energy loss. 

\begin{figure*}[!tbhp]
\begin{center}
\epsfxsize 6.5cm
\epsffile {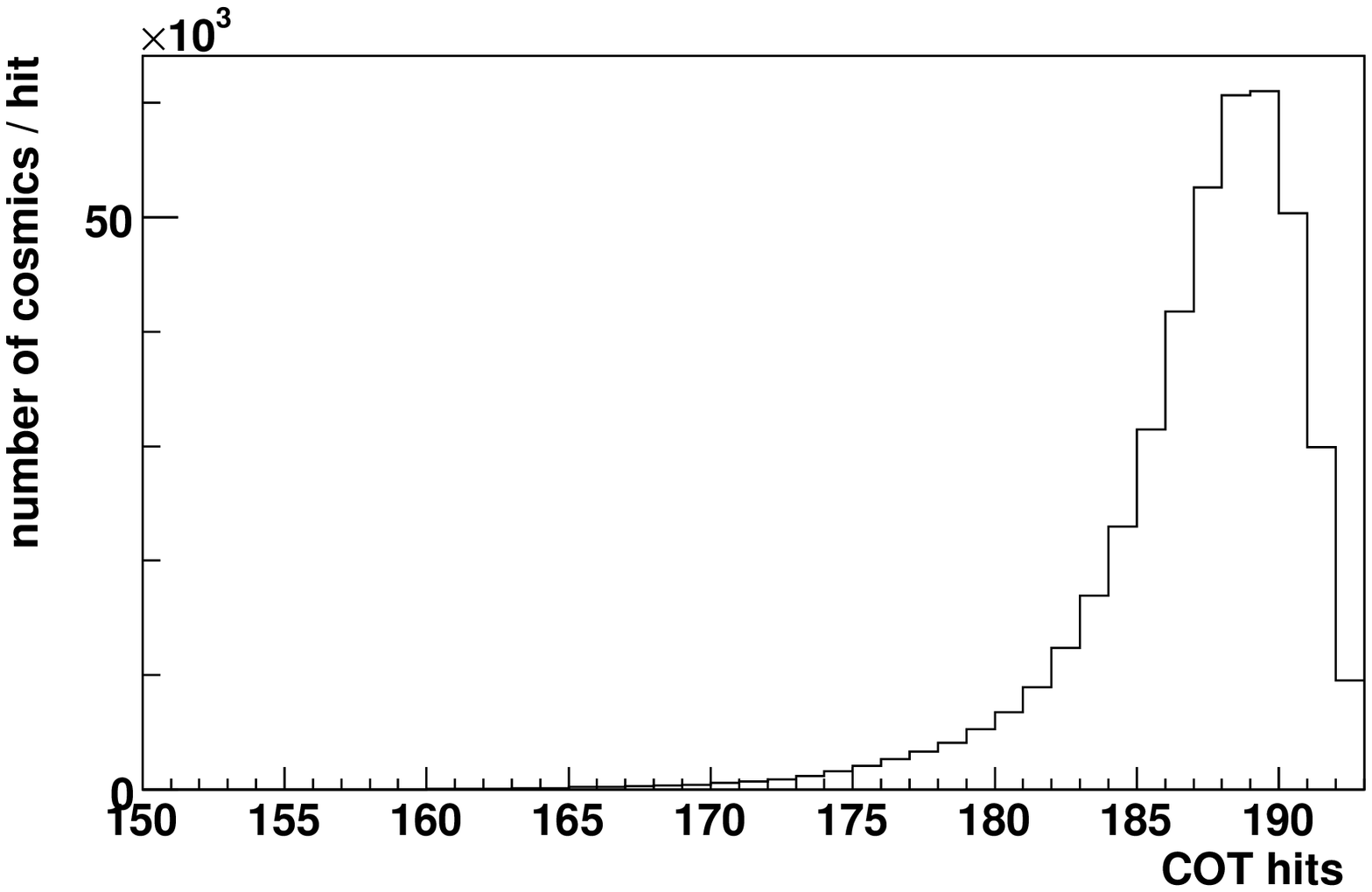}
\epsfxsize 6.5cm
\epsffile {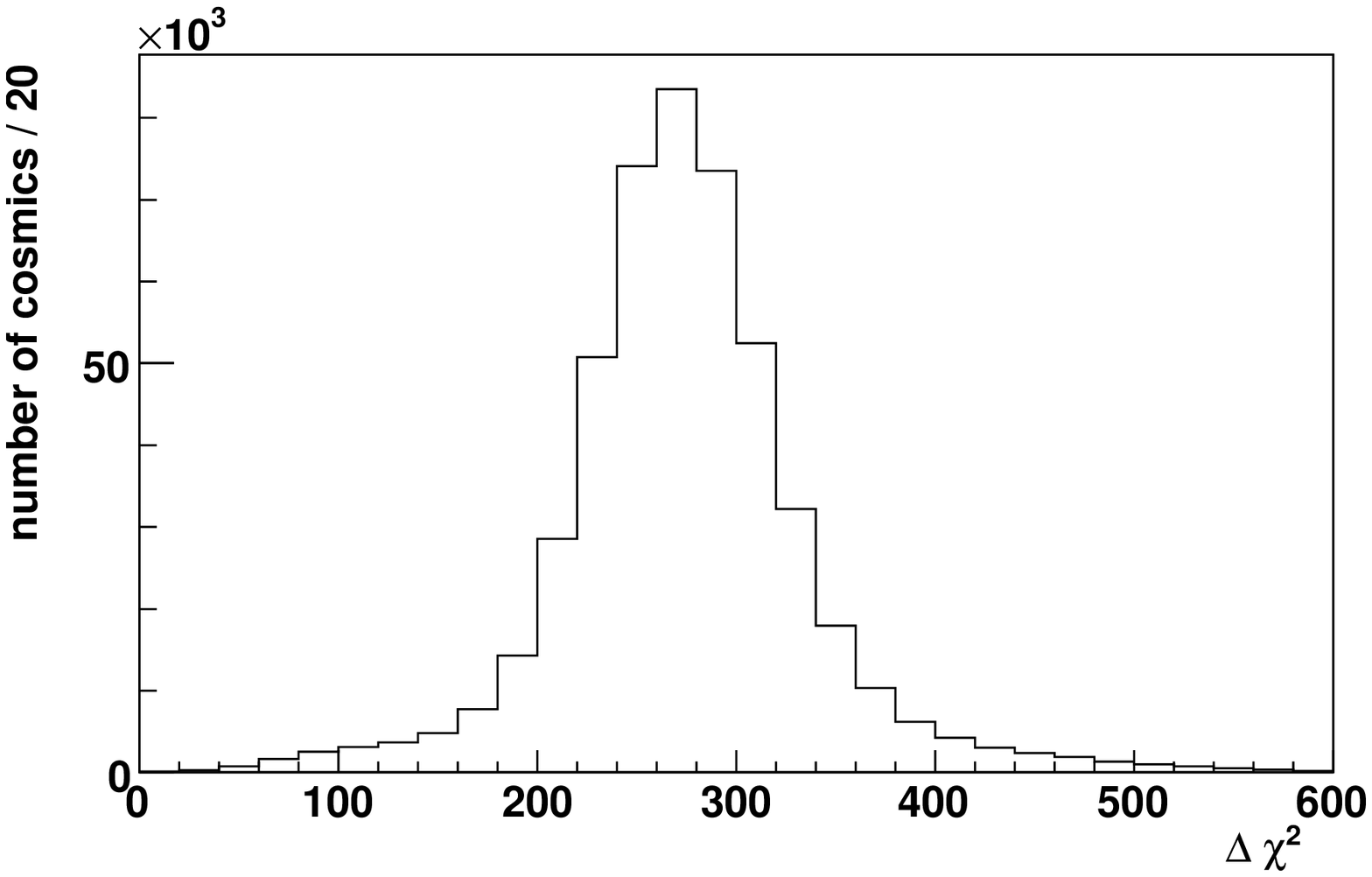}
\end{center}
\caption{Distributions of (left) the number of COT hits on the dicosmic track  and (right) the difference in $\chi^2$ between the best-fit and next-best-fit hypotheses for the directions of propagation of the two track segments.   }
\label{fig:dicosmicNhits}
\end{figure*}

As described in~\cite{cosmicNIM}, the dicosmic track is seeded by a muon track found by the COT track reconstruction software which is used for collider tracks. We eliminate seed tracks with
 $p_T < 10$~GeV to protect against potential problems with the triggering and 
 reconstruction of out-of-time cosmic rays. After seeding and associating hits to the dicosmic track, 
 the dicosmic track fit~\cite{cosmicNIM} allows the direction of propagation
 and the time of incidence of the cosmic ray to be free parameters. To ensure full containment in the COT, the dicosmic track is required to have hits in the outermost COT superlayer at both ends
 of its trajectory. As shown in Fig.~\ref{fig:dicosmicNhits}, the efficiency of associating hits (from a maximum possible 192 hits) with the dicosmic track 
 is $\approx 97$\% for this sample. In addition to the cosmic-ray hypothesis, two additional hypotheses
 are also tested in the dicosmic fit: a pair of outgoing back-to-back tracks (as could be produced from a two-body decay at rest), and a pair of incoming tracks (indicating a reconstruction problem in one of the
 two track segments). We require the cosmic-ray hypothesis to be the $\chi^2$-minimizing hypothesis. The difference $\Delta \chi^2$ from the next most-likely hypothesis is an indicator of the robustness
 of the dicosmic track fit. The distribution of $\Delta \chi^2$, shown in Fig.~\ref{fig:dicosmicNhits}, indicates a very strong preference for the cosmic-ray hypothesis in our selected sample.   Finally, we 
 require $|z_0| < 60$~cm to emulate the collider-track sample to which the resulting alignment constants will be applied, since 
 CDF physics analyses make this requirement to select tracks from the
 longitudinal beam-crossing region.  

\section{Endplate alignment corrections}
\label{sec:Corrections}
\begin{figure*}[!tbhp]
\begin{center}
\epsfxsize 15cm
\epsffile {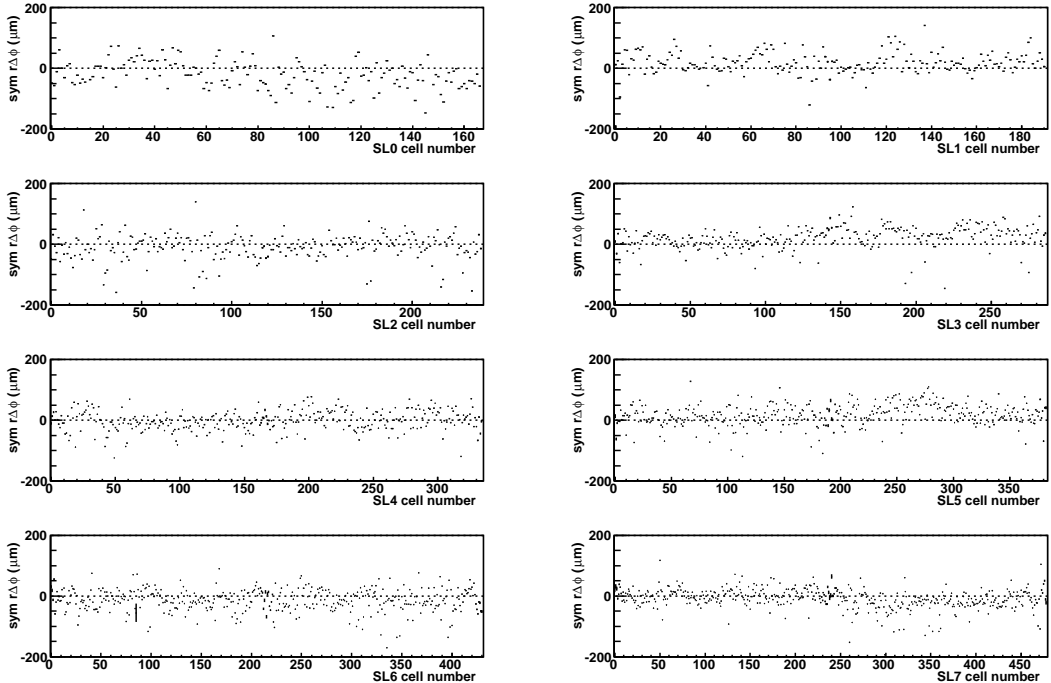}
\end{center}
\caption{Symmetric alignment corrections measured with cosmic-ray residuals, after the CMM measurements are applied in the track reconstruction. The superlayers are numbered starting from
 the innermost ({\sc sl0}) to the outermost ({\sc sl7}). The plots on the left (right) correspond to the stereo (axial) superlayers.   }
\label{fig:align_a_cell_before}
\end{figure*}

\begin{figure*}[!tbhp]
\begin{center}
\epsfxsize 15cm
\epsffile {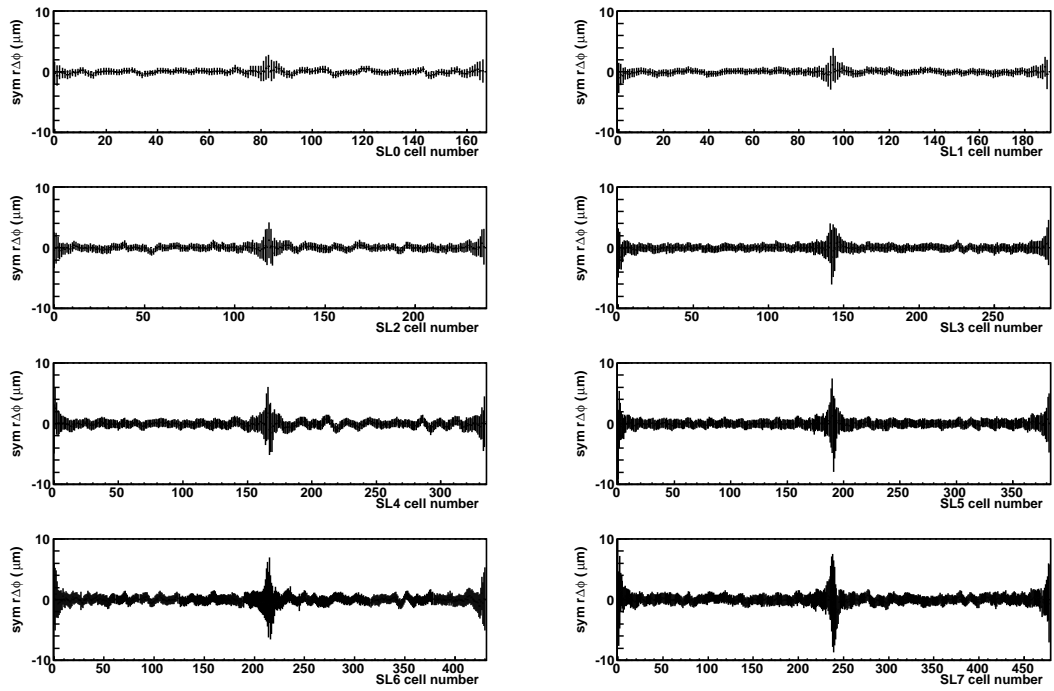}
\end{center}
\caption{Symmetric alignment constant residuals after applying the cosmic-ray corrections. Note that the vertical scale is smaller by a factor of 20 compared to Fig.~\ref{fig:align_a_cell_before}.
    }
\label{fig:align_a_cell_after}
\end{figure*}

 The starting point for obtaining the alignment correction constants is the cell-position measurements~\cite{cot}  made 
 using a coordinate-measuring machine (CMM). The cell-tilt angles were updated using a calculation~\cite{fea} based on a finite-element analysis (FEA) which predicted a cell-tilt correction
  of 1.5 mrad\footnote{An independent cross-check conducted with the cosmic rays confirmed this prediction within 25\%. }.  These measurements are used as input for the reconstruction of
 the cosmic-ray data. 
 The symmetric alignment correction constants extracted from these data for each cell are shown for all superlayers in Fig.~\ref{fig:align_a_cell_before}.  
 The magnitude of these corrections is consistent with the nominal accuracy of the CMM measurements. After applying 
 the alignment corrections iteratively, the residuals converge to zero within a statistical precision of $\approx 0.5~\mu$m, as shown in Fig.~\ref{fig:align_a_cell_after}. 

\begin{figure*}[!tbhp]
\begin{center}
\epsfxsize 15cm
\epsffile {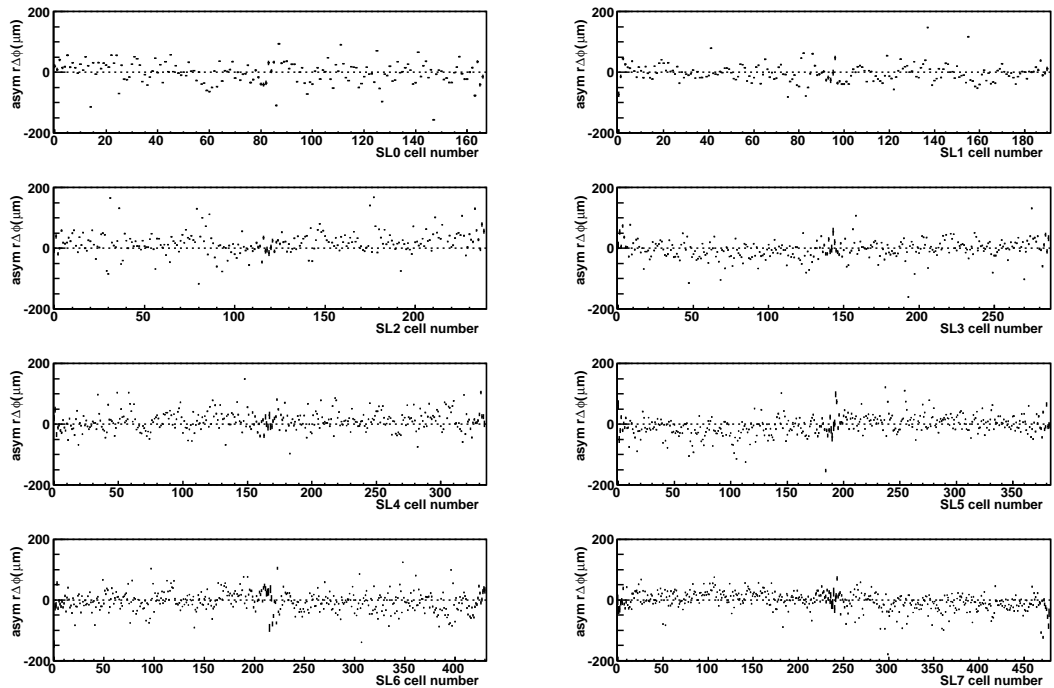}
\end{center}
\caption{Asymmetric alignment corrections measured with cosmic-ray residuals, after the CMM measurements are applied in the track reconstruction.   }
\label{fig:align_c_cell_before}
\end{figure*}

\begin{figure*}[!tbhp]
\begin{center}
\epsfxsize 15cm
\epsffile {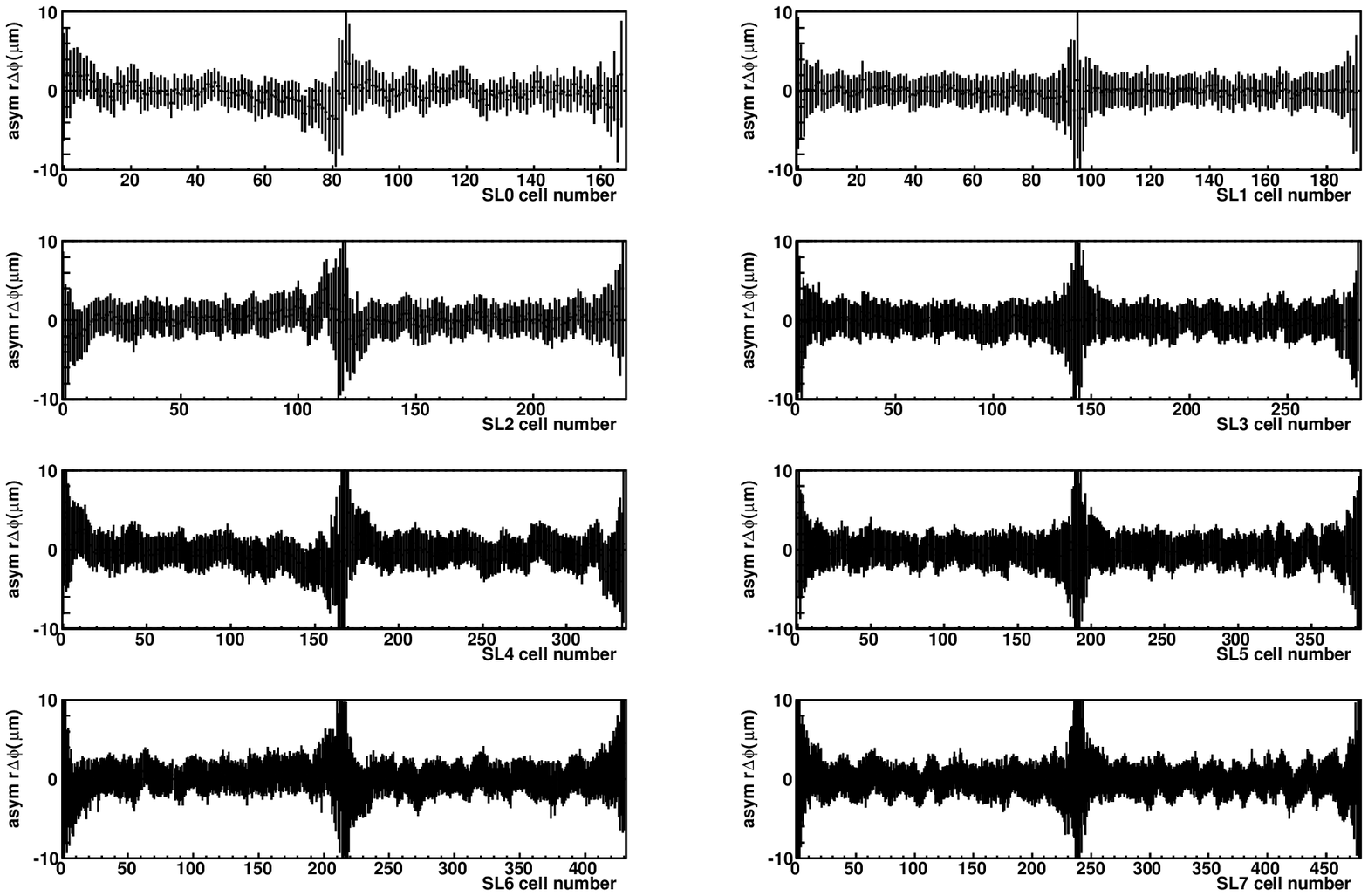}
\end{center}
\caption{Asymmetric alignment residuals after applying the cosmic-ray corrections. Note that the vertical scale is smaller by a factor of 20 compared to Fig.~\ref{fig:align_c_cell_before}.  }
\label{fig:align_c_cell_after}
\end{figure*}

 The asymmetric alignment correction constants extracted from the cosmic-ray data are shown for each cell in all superlayers 
 in Fig.~\ref{fig:align_c_cell_before}. 
 These constants are defined as half of the difference between the east and west endplate corrections. The $| z_0 | < 60$~cm requirement not only reduces the 
 total cosmic-ray sample size by a factor of 2.5, but also reduces
 the lever-arm for measuring the $z$-dependence of the residuals, particularly for the superlayers at small radius. As a result, 
 the asymmetric constants have a statistical precision of $\approx 2~\mu$m. After applying these corrections iteratively, the mean residuals shown in Fig.~\ref{fig:align_c_cell_after} 
 are obtained. The convergence to zero is very good, except for a few cells that are close to the horizontal plane\footnote{The horizontal plane is defined by $\phi = 0$ where cells 
 numbered zero are located. Cells numbered half of the maximum are located at $\phi = \pi$. } 
 in the upper half of the stereo superlayers (labelled {\sc sl0, sl2, sl4} and {\sc sl6} respectively). 

\begin{figure*}[!tbhp]
\begin{center}
\epsfxsize 15cm
\epsffile {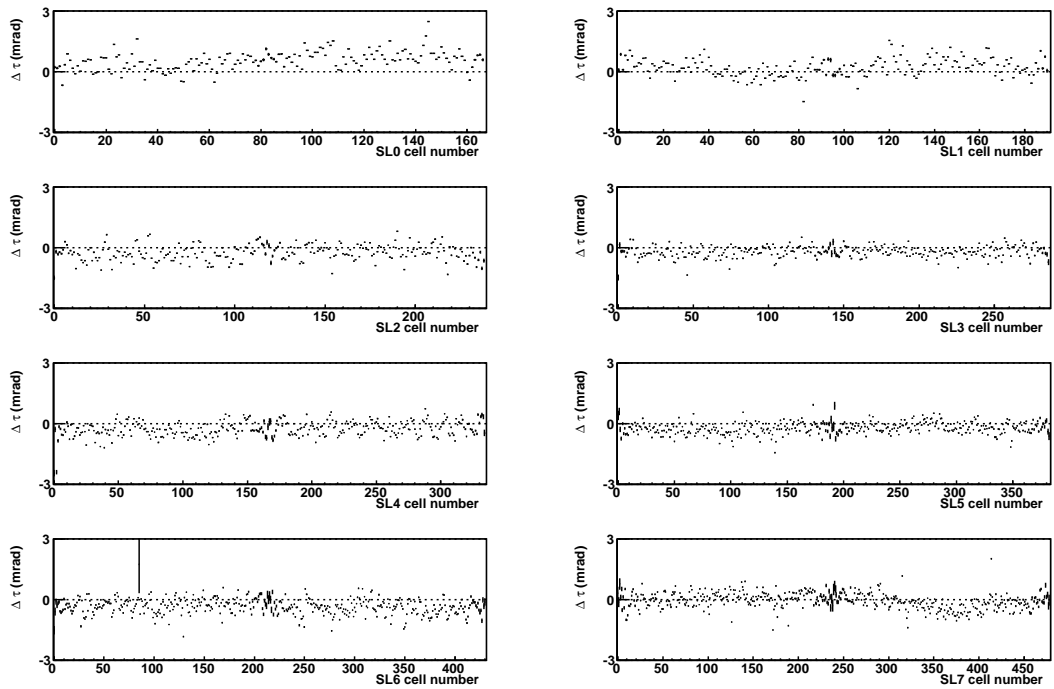}
\end{center}
\caption{Cell-tilt angle corrections measured with cosmic-ray residuals, after the CMM measurements are applied in the track reconstruction.   }
\label{fig:align_b_cell_before}
\end{figure*}

\begin{figure*}[!tbhp]
\begin{center}
\epsfxsize 15cm
\epsffile {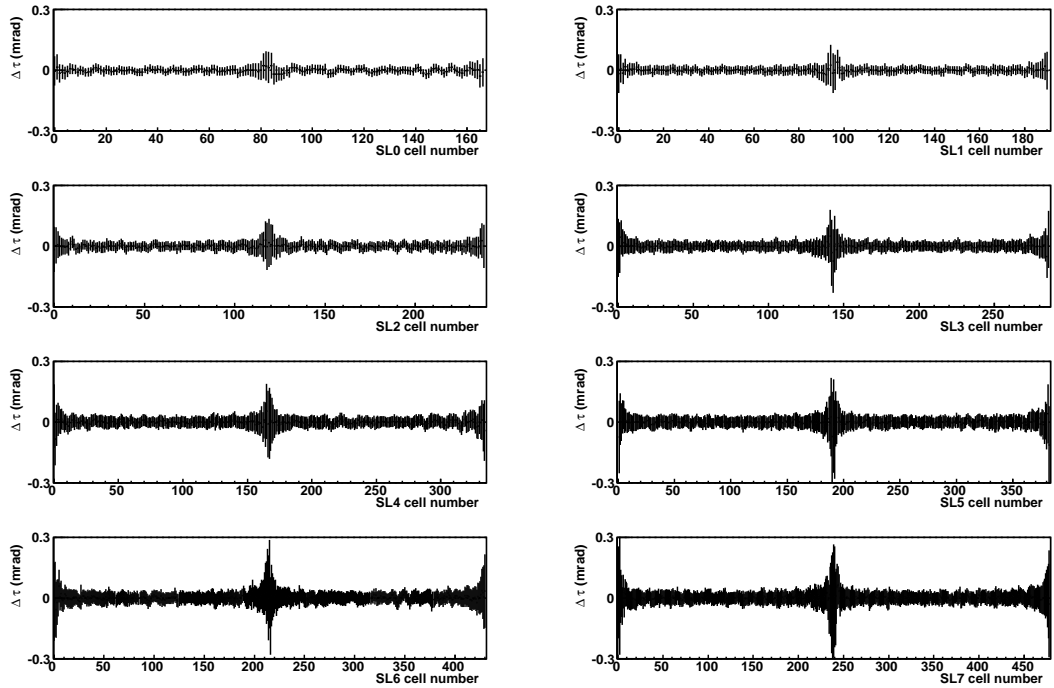}
\end{center}
\caption{Cell-tilt angle residuals after applying the cosmic-ray corrections. Note that the vertical scale is smaller by a factor of 10 compared to Fig.~\ref{fig:align_b_cell_before}.  }
\label{fig:align_b_cell_after}
\end{figure*}

 The corrections to the cell-tilt angle before applying the cosmic-ray alignment procedure are shown in 
 Fig.~\ref{fig:align_b_cell_before}; the residual deviations after a few iterations are shown in Fig.~\ref{fig:align_b_cell_after}. The corrections are $\approx 500~\mu$rad and 
 the deviations converge to zero within a typical statistical precision of $\approx 20~\mu$rad. 
\section{Gravitational sag and electrostatic deflection}
\label{sec:wireShape}
The corrections discussed above determine the wire positions at the endplates. 
 The reconstructed hit coordinates also depend on the shape of the wires between the endplates. Transverse forces on the 3 m-long wires due to gravity
 and electrostatics cause the wires to deviate from straight lines by {\it O}(200~$\mu$m). 
 To first order, the gravitational sag acts in the vertical direction and the electrostatic deflections
 occur  in the direction perpendicular to the cell tilt angle ($35\,^{\circ}$ at
 $\phi = 0$). The field sheets also experience similar forces and resulting deflections. The electrostatic deflection depends on the distance of the wire
 from the two field sheets, coupling the deviations due to gravity and electrostatics in an azimuth-dependent way. The deflections also depend on the tension 
 in the wires. The wire
 shape is modeled as a combination of a  $V$-shape and a parabola as functions of $z$, to include the weight of spacers in the middle of the wires. 
 The maximum magnitude of these functions, defined to occur at $z=0$, is varied 
 as a function of
 azimuth and radius as discussed below. 

 In the CDF track reconstruction, the vertical  deflection ($\delta y$) of the wires with respect to their positions at $z = \pm 155$~cm due to gravitational sag
 is parameterized as follows~\cite{cot}:
\begin{equation}
- \delta y ~ (\mu {\rm m}) = 59 ( 1 - |Z|) + 203 (1 - Z^2 ) 
\label{eqn:gravityZshape}
\end{equation}
 where $Z \equiv z/(155~{\rm cm})$ and $z$ is the longitudinal coordinate along the wire.  This gravitational sag function is illustrated in Fig.~\ref{fig:zShapes}. 
 As discussed in~\cite{cot}, the coefficients were calculated using the
 median linear density of the wire and its nominal tension, which are expected to vary by $\pm 3$\% and $\pm 5$\% respectively. 
 The first term corresponds to the center support and the second term corresponds to the weight of the wire. 
 
 The deflection due to electrostatic forces ($\xi$) is parameterized with the same $z$-dependent function but in the direction perpendicular to the sense wire plane, 
\begin{equation}
\xi = m(\phi) \left[ 0.29 ( 1 - |Z|) + (1 - Z^2 ) \right].
\label{origZwireShape}
\end{equation}
The coefficient $0.29 = 59 / 203$ is the ratio of the two terms derived from Eqn.~\ref{eqn:gravityZshape}, such that we use the same wire shape as a function of $z$ for describing both
 gravitational sag and electrostatic deflection.
 The  magnitude modulation function $m(\phi)$ is measured~\cite{cot} to vary sinusoidally with the azimuthal orientation $\phi_{\rm wp}$ of the wire plane, 
\begin{equation}
m (\phi) =  a \cos \phi_{\rm wp} + o
\label{wireAmplitude}
\end{equation}
 where the azimuthal dependence arises from the different gravitational sags of the sense wires and the field sheets. 
  The sign and phase convention is such that the electrostatic deflection of the sense wire plane is upwards (opposite the gravitational sag) when  the sense wire plane
 is horizontal. The amplitude $a = 117~\mu$m  and the offset $o = -84~\mu$m  were measured in a preproduction prototype with an electric field of 2.4 kV/cm, 
 and scaled down by a factor of 2.7
 to account for the COT operating point of 1.9 kV/cm~\cite{cot}, such that $a = 43~\mu$m and $o = -31~\mu$m are applied to the collider data. 

 The alignment corrections and the wire-shape functions
 can be cross-checked by comparing the track parameters of the two segments of the cosmic ray track. The sum or difference (depending on the definition 
 of the track parameter) of the track parameters, defined at the point of closest approach to the beamline, should be consistent with zero within resolution. 
 The convention for computing the pull for $\phi_0$, $z_0$ and $t_0$ is that the outgoing segment is positive and incoming segment is negative. In the case of
 curvature, $d_0$ and $\cot \theta$, the pull is defined as the sum of the two segments' parameters, since the sign changes from top to bottom. 
 
\begin{figure*}[!tbhp]
\begin{center}
\epsfxsize 15cm
\epsffile {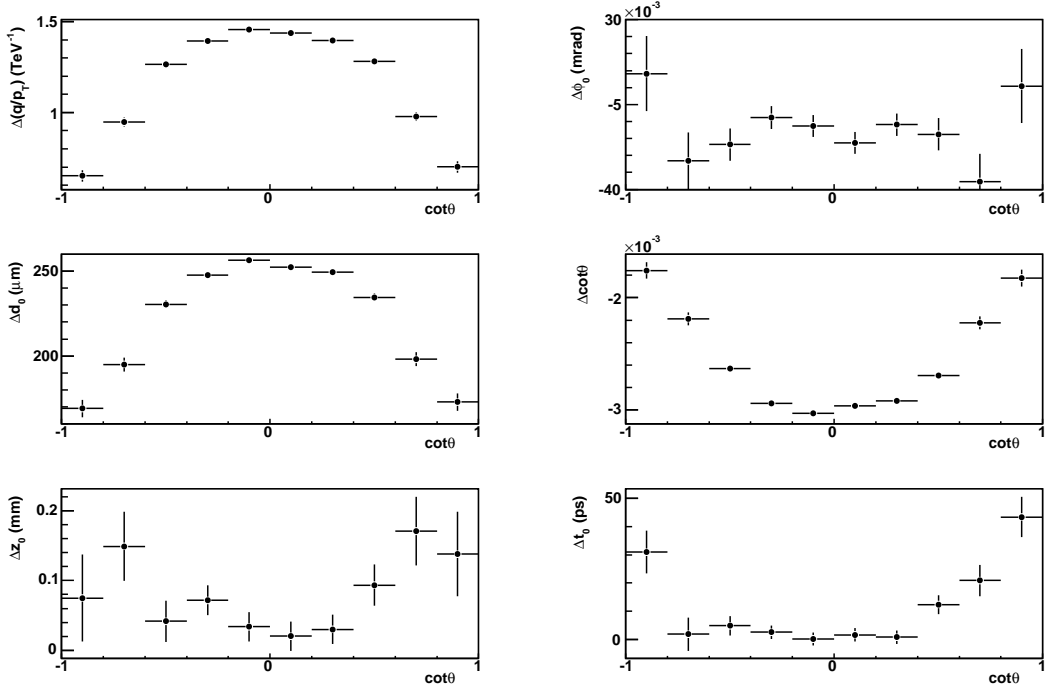}
\end{center}
\caption{Track parameter pulls as a function of $\cot \theta$ after the cosmic-ray alignment corrections and the default wire-shape functions
 are applied in the track reconstruction.   }
\label{fig:pullsVsCotTheta_afterNoR}
\end{figure*}

\begin{figure*}[!tbhp]
\begin{center}
\epsfxsize 14cm
\epsffile {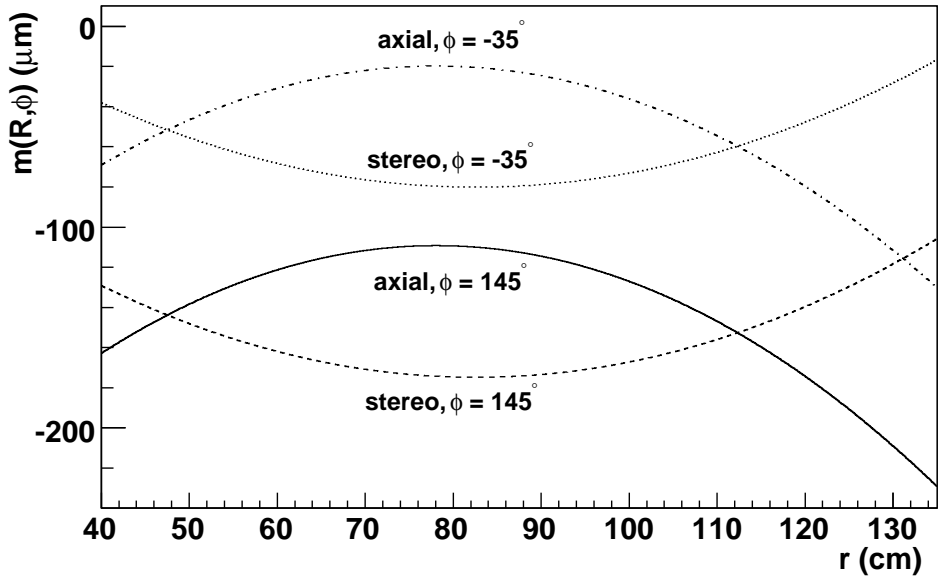}
\end{center}
\caption{The radius-dependent electrostatic deflection at $z=0$, for axial and stereo superlayers. The sinusoidal dependence on azimuth is bounded by the curves shown for 
 $\phi = -35^{\circ}$ and $\phi = 145^{\circ}$ respectively, where $\cos \phi_{\rm wp} = \pm 1$ in Eqn.~\ref{wireAmplitudeNew}.    }
\label{fig:radial}
\end{figure*}

\begin{figure*}[!tbhp]
\begin{center}
\epsfxsize 13.0cm
\epsffile {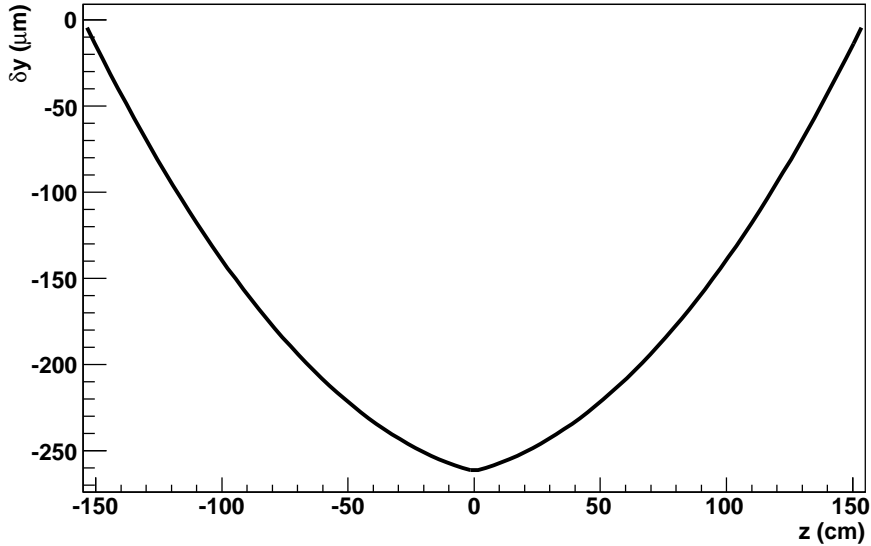}
\end{center}
\caption{The gravitational sag of the wires (Eqn.~\ref{eqn:gravityZshape})
  as a function of $z$. The same $z$-dependence,  normalized by the magnitude modulation function $m(R, \phi)$ (Eqns.~\ref{origZwireShape} and~\ref{wireAmplitudeNew}) also describes 
the electrostatic deflection perpendicular to the sense wire plane.    }
\label{fig:zShapes}
\end{figure*}

 Our studies of the track-parameter pulls in the cosmic-ray data show significant dependence on $\cot \theta$ after the above wire-shape functions are
 applied, as shown in Fig.~\ref{fig:pullsVsCotTheta_afterNoR}. In particular, the mean pulls for the curvature, impact parameter and $\cot \theta$ show a large
 quadratic dependence on $\cot \theta$. 
 We find that these variations of the pulls can be reduced by incorporating a radius-dependent magnitude modulation function $m(\phi) \to m(R, \phi)$, as follows,
\begin{equation}
m (R, \phi) =  a(R) \cos \phi_{\rm wp} + o(R),
\label{wireAmplitudeNew}
\end{equation}
 where  $R \equiv r/(140~{\rm cm})$ and $r$ is the radial position of the wire. 
 We use the pull measurements from the cosmic-ray data to tune the radius-dependent offset function, $o(R) = o + \delta o(R)$, where  
\begin{equation}
 \delta o(R) =  b_0 + b_1 R + b_2 R^2 \; \; .
\label{offsetFunction}
\end{equation}
 The coefficients $b_{i}$ account for the inward bending of the endplates 
 as a function of  radius (see Fig.~10 of~\cite{cot})  and the possible effect on the wire tension, resulting in radius-dependent corrections to the electrostatic deflection. 
 The stereo superlayers experience additional electrostatic effects due to the twisting forces induced by the stereo angle~\cite{cot}. 
 Therefore, we tune the  coefficients in Eqn.~\ref{offsetFunction} separately for the axial and stereo superlayers in order to minimize the remaining biases, with the resulting values shown in
 Table~\ref{wireShapes}. A small correction to the sinusoidal amplitude $a(R)$ is also introduced
 to reduce pull biases as a function of $\phi_0$:
\begin{equation}
 a(R) =  a \left[ 1 + 0.03 \frac{\delta o(R)}{ o} \right] \; \; ,
\label{amplitudeFunction}
\end{equation}
 resulting in $a(R)$ varying between 45~$\mu$m and 49~$\mu$m for the different superlayers. The values of $o(R)$ and $a(R)$ that we use are shown in Table~\ref{wireShapesNew}.
The values of $o(R)$ are modified substantially compared to the nominal value of $-31~\mu$m, justifying the inclusion of the additional degrees of freedom in Eqn.~\ref{offsetFunction}.
  The radial dependence of the wire shapes deduced from the cosmic-ray data is illustrated in 
 Fig.~\ref{fig:radial}. 
  
 The replacement of $m(\phi)$ by $m(R, \phi)$ substantially reduces the dependence of the pulls on $\cot \theta$, $z_0$, 
and $\phi_0$, as shown in Sec.~\ref{sec:Biases}. 
\begin{table}
\begin{center}
\begin{tabular}{l|ccc}
\hline
\hline
         & $b_0$ & $b_1$ & $b_2$ \\
\hline
axial superlayers & -250 & 777 & -696 \\
\hline
stereo superlayers    &  68  & -558 & 473  \\
\hline
\end{tabular}
\caption{Parameters in the wire-shape functions described in Eqn.~\ref{offsetFunction}, in units of $\mu$m.  }
\label{wireShapes}
\hspace*{1cm}
\end{center}
\end{table}
\begin{table}
\begin{center}
\begin{tabular}{l|cc}
\hline
\hline
superlayer         & $a(R)$~($\mu$m) & $o(R)$~($\mu$m)  \\
\hline
{\sc sl0} & 46 & -96 \\
\hline
{\sc sl1} & 45 & -78 \\
\hline
{\sc sl2} & 47  & -124 \\
\hline
{\sc sl3} & 45 & -65 \\
\hline
{\sc sl4} & 47  & -124  \\
\hline
{\sc sl5} & 46 & -91 \\
\hline
{\sc sl6} & 46 & -98 \\
\hline
{\sc sl7} & 49 & -157 \\
\hline
\end{tabular}
\caption{Parameters describing the electrostatic deflection varying with azimuth as given by Eqn.~\ref{wireAmplitudeNew}.  }
\label{wireShapesNew}
\hspace*{1cm}
\end{center}
\end{table}
\section{Drift model}
\label{sec:driftModel}
The wire positions are transferred to  hit positions by the drift model,  which converts the drift-time measurement to a hit distance from the wire. The local coordinate is denoted by $Y$
 (where the sign denotes whether the coordinate is on the left or right side of the sense wire in the cell) and the local distance is denoted by $D \equiv |Y|$. The  residual
 $\Delta Y \equiv Y_{\rm hit} - Y_{\rm track}$ is used for the alignment results discussed above. The residual
  $\Delta D \equiv  {\rm sign}(Y_{\rm track}) \times \Delta Y$,  is used to study
 the drift model. The simplest drift model would be parameterized by a timing offset (due to propagation delays in the electronic read-out path) and a drift speed,  if the time-distance relationship
 were proportional. In practice, this simple drift model describes the time-distance relationship well in the $\approx$~2-7 mm range of drift distance. The regions close to the wire and
 the field sheets require additional parameters to describe the non-linear time-distance relationships in those regions. An elaborate drift model has been developed and is used in the CDF track
 reconstruction~\cite{cot}. 

\begin{figure*}[!tbhp]
\begin{center}
\epsfxsize 15cm
\epsffile {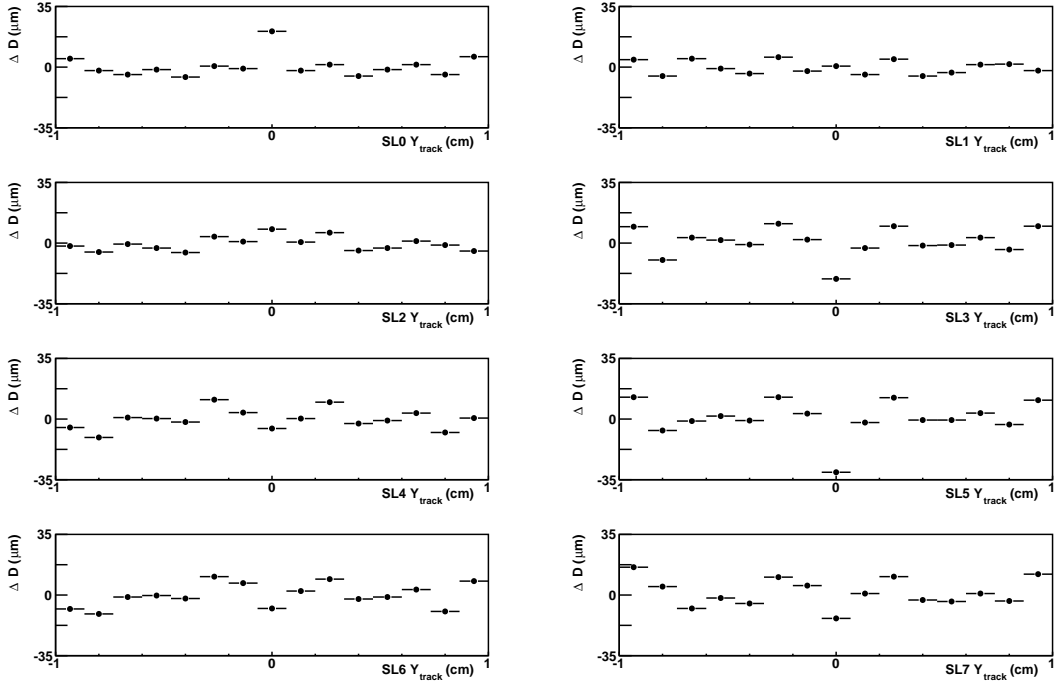}
\end{center}
\caption{The mean $\Delta D$ as a function of local track coordinate $Y_{\rm track}$ in the cell for the eight superlayers.   }
\label{fig:diCosmicDeltaDvsTrackY}
\end{figure*}

 For the cosmic-ray-based alignment study discussed here, we perform additional tuning of this drift model, particularly in the non-linear regions, by iteratively
 reducing the dependence of $\Delta D$ on $Y_{\rm track}$. The final results for each superlayer are shown
  in Fig.~\ref{fig:diCosmicDeltaDvsTrackY}.  In addition to tuning the drift model parameters, we
 also tune the timing offset for each cell so that $\langle \Delta D \rangle = 0 $ over all hits in the cell. The differences in timing offsets for different wires within a cell, as well
 as their variation with time,  have been separately
 calibrated by an electronic calibration pulse injection system~\cite{cot}.  As Fig.~\ref{fig:diCosmicDeltaDvsTrackY} shows, the residuals $\Delta D$ are typically less than 10~$\mu$m over most of the
 drift region. The remaining mismodeling in the nearest and farthest $\approx 1$~mm region from the wires, of about 20~$\mu$m, is  small compared to the hit resolution of $\approx 140~\mu$m. 
\section{Cross-checks of track parameter biases}
\label{sec:Biases}
\begin{figure*}[!tbhp]
\begin{center}
\epsfxsize 15cm
\epsffile {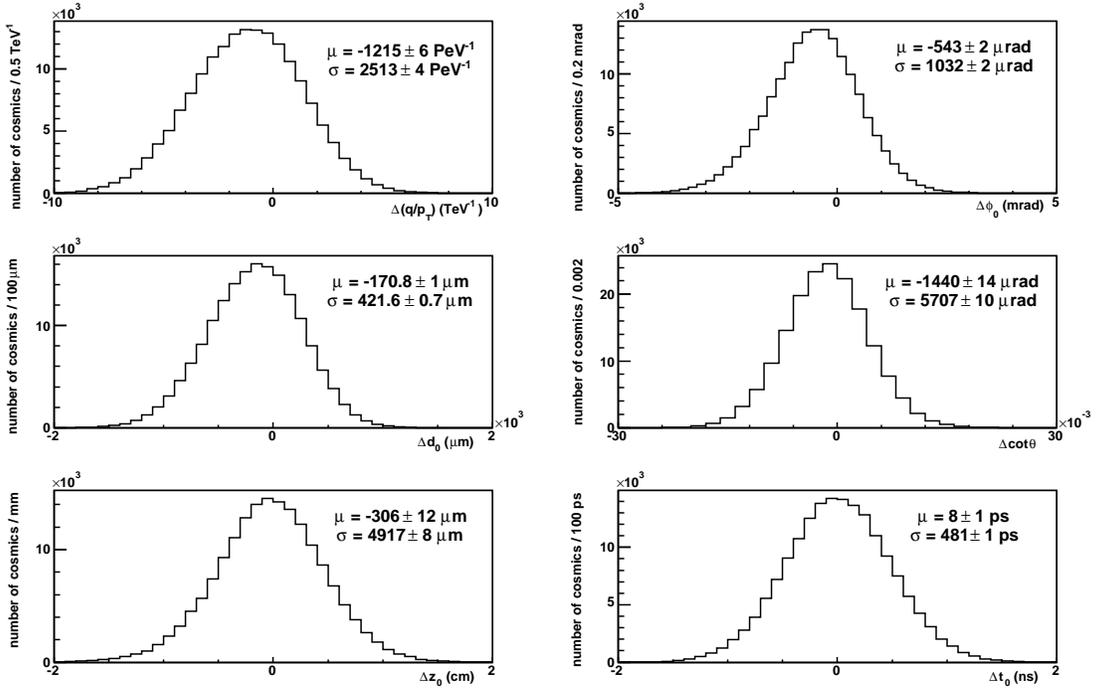}
\end{center}
\caption{Track parameter pulls after the CMM measurements are applied in the track reconstruction, shown with the mean ($\mu$) and standard deviation ($\sigma$).   }
\label{fig:pulls_before}
\end{figure*}

\begin{figure*}[!tbhp]
\begin{center}
\epsfxsize 15cm
\epsffile {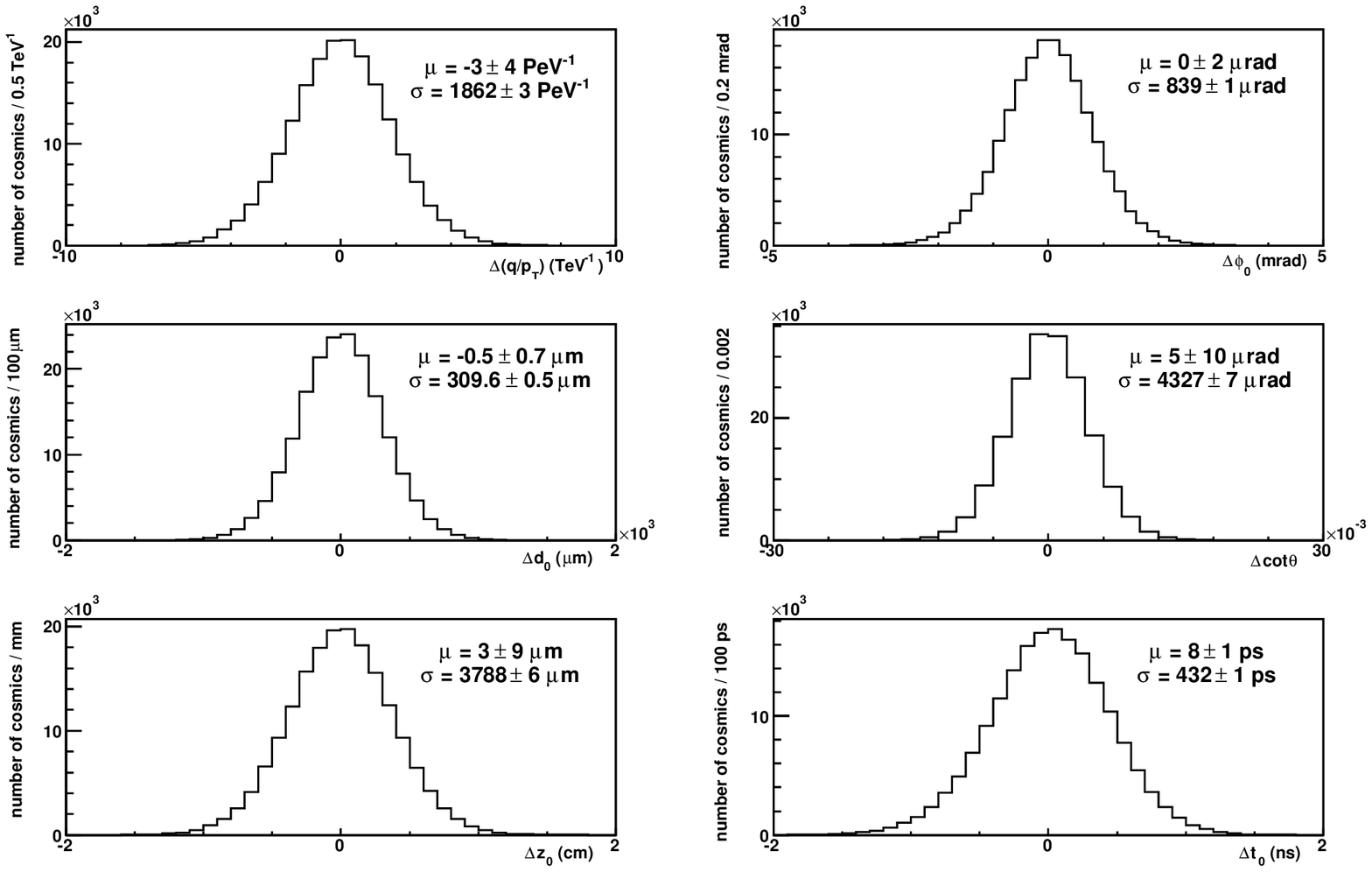}
\end{center}
\caption{Track parameter pulls after applying the cosmic-ray corrections, shown with the mean ($\mu$) and standard deviation ($\sigma$).   }
\label{fig:pulls_after}
\end{figure*}

The distributions 
 of the pulls 
 are shown in Fig.~\ref{fig:pulls_before} using the CMM measurements and in Fig.~\ref{fig:pulls_after} after the cosmic-ray-based alignment corrections and our wire-shape tunes 
 are applied. The average
 values are consistent with zero, and the resolution is improved by $\approx 25$\%, after the alignment. 

\begin{figure*}[!tbhp]
\begin{center}
\epsfxsize 15cm
\epsffile {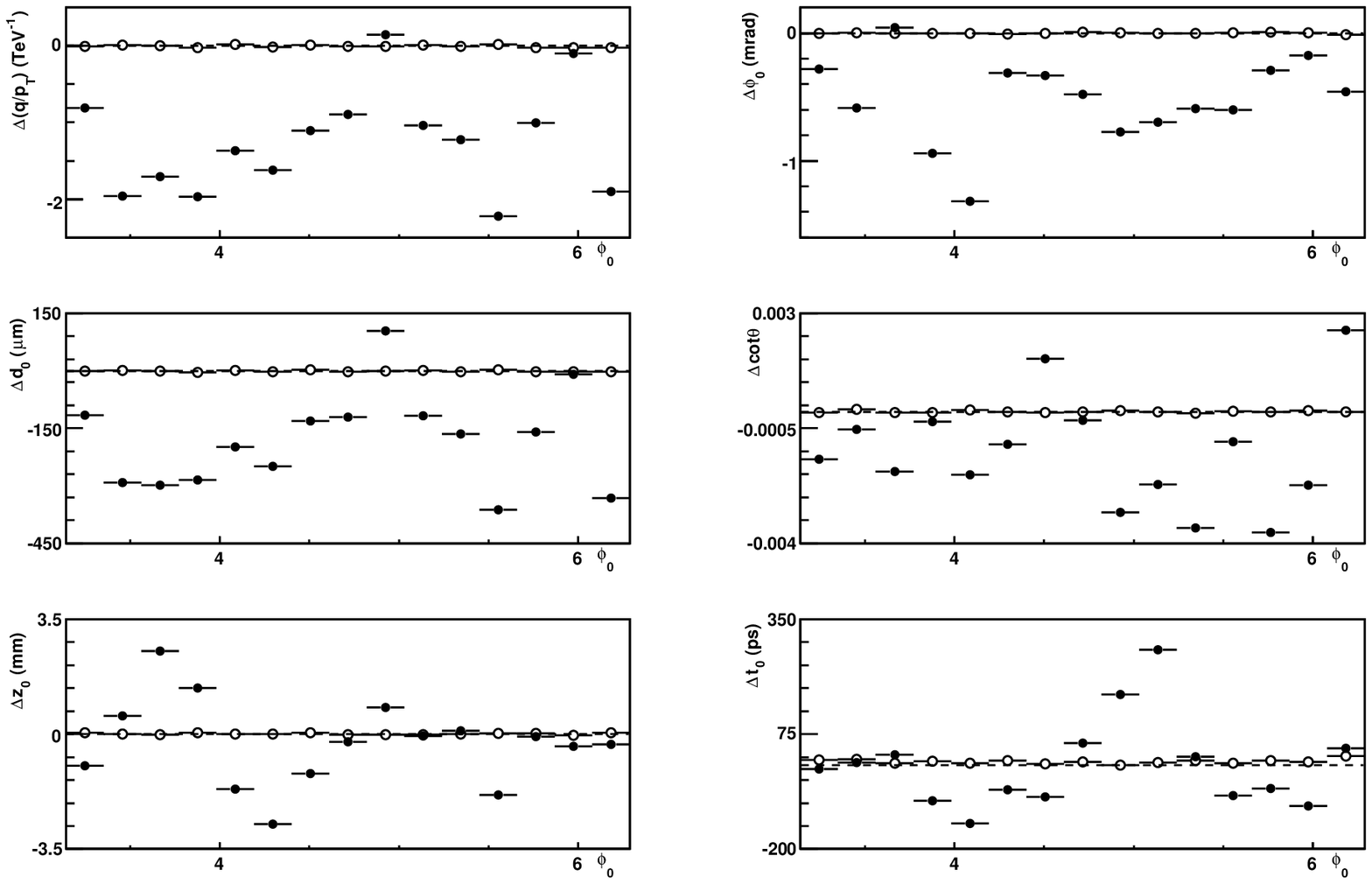}
\end{center}
\caption{Track parameter pulls as functions of $\phi_0$ before (solid circles) and after (open circles) the cosmic-ray alignment.   }
\label{fig:pullsVsPhi0Overlay}
\end{figure*}

\begin{figure*}[!tbhp]
\begin{center}
\epsfxsize 15cm
\epsffile {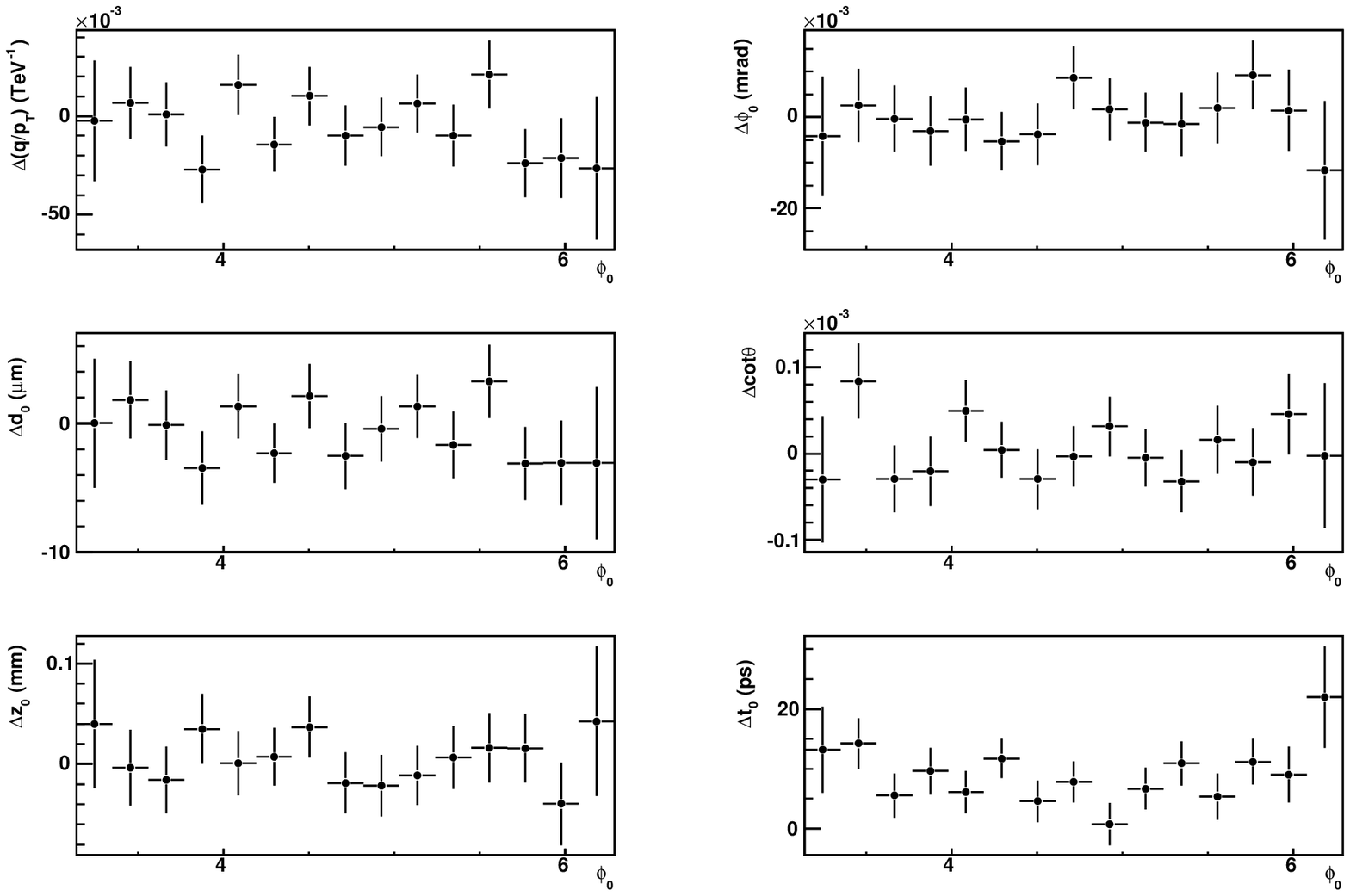}
\end{center}
\caption{Track parameter pulls as functions of $\phi_0$ after applying the cosmic-ray corrections.   }
\label{fig:pullsVsPhi0_after}
\end{figure*}

\begin{figure*}[!tbhp]
\begin{center}
\epsfxsize 15cm
\epsffile {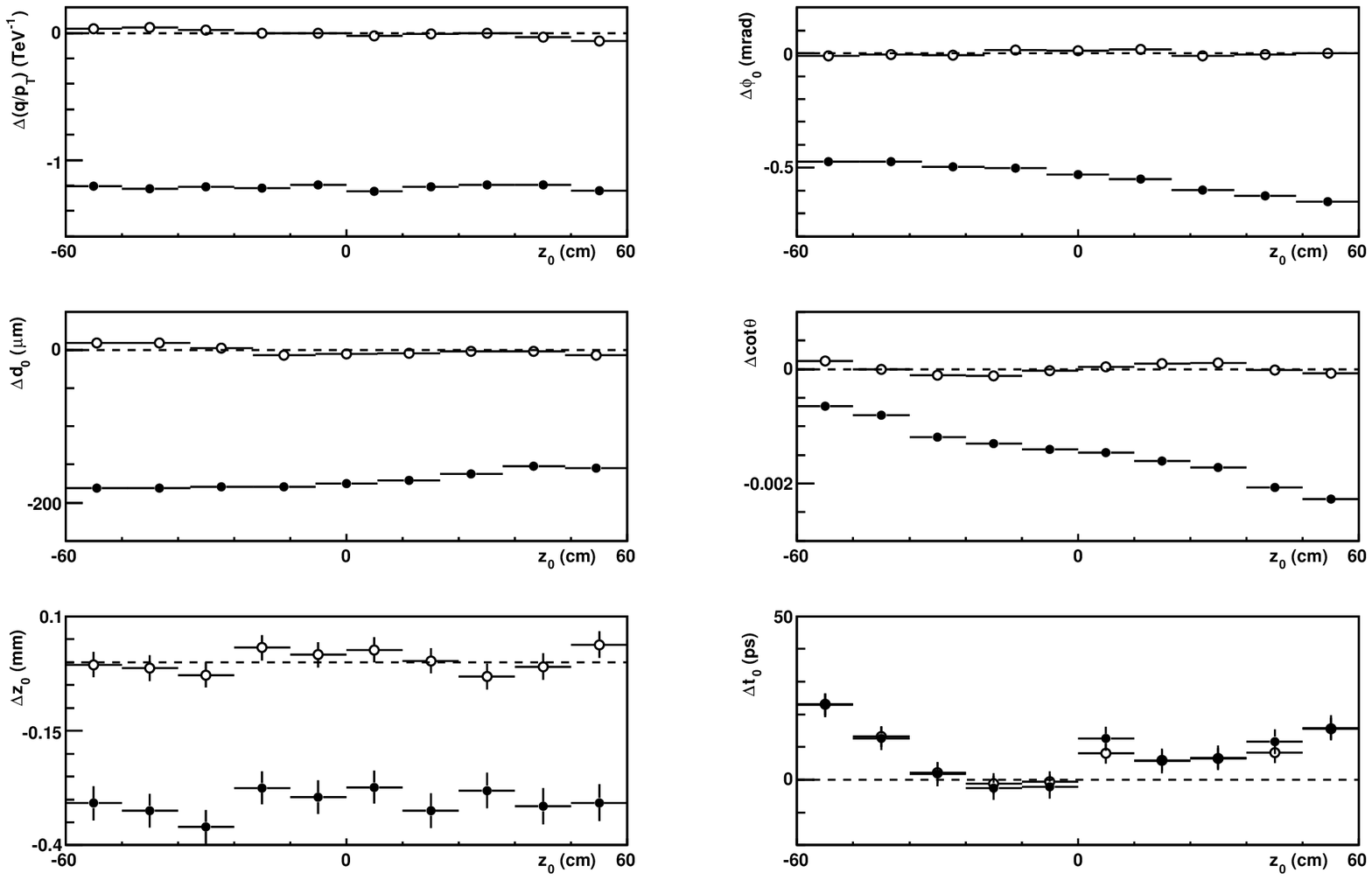}
\end{center}
\caption{Track parameter pulls as functions of $z_0$ before (solid circles) and after (open circles) the cosmic-ray alignment.   }
\label{fig:pullsVsZ0Overlay}
\end{figure*}

\begin{figure*}[!tbhp]
\begin{center}
\epsfxsize 15cm
\epsffile {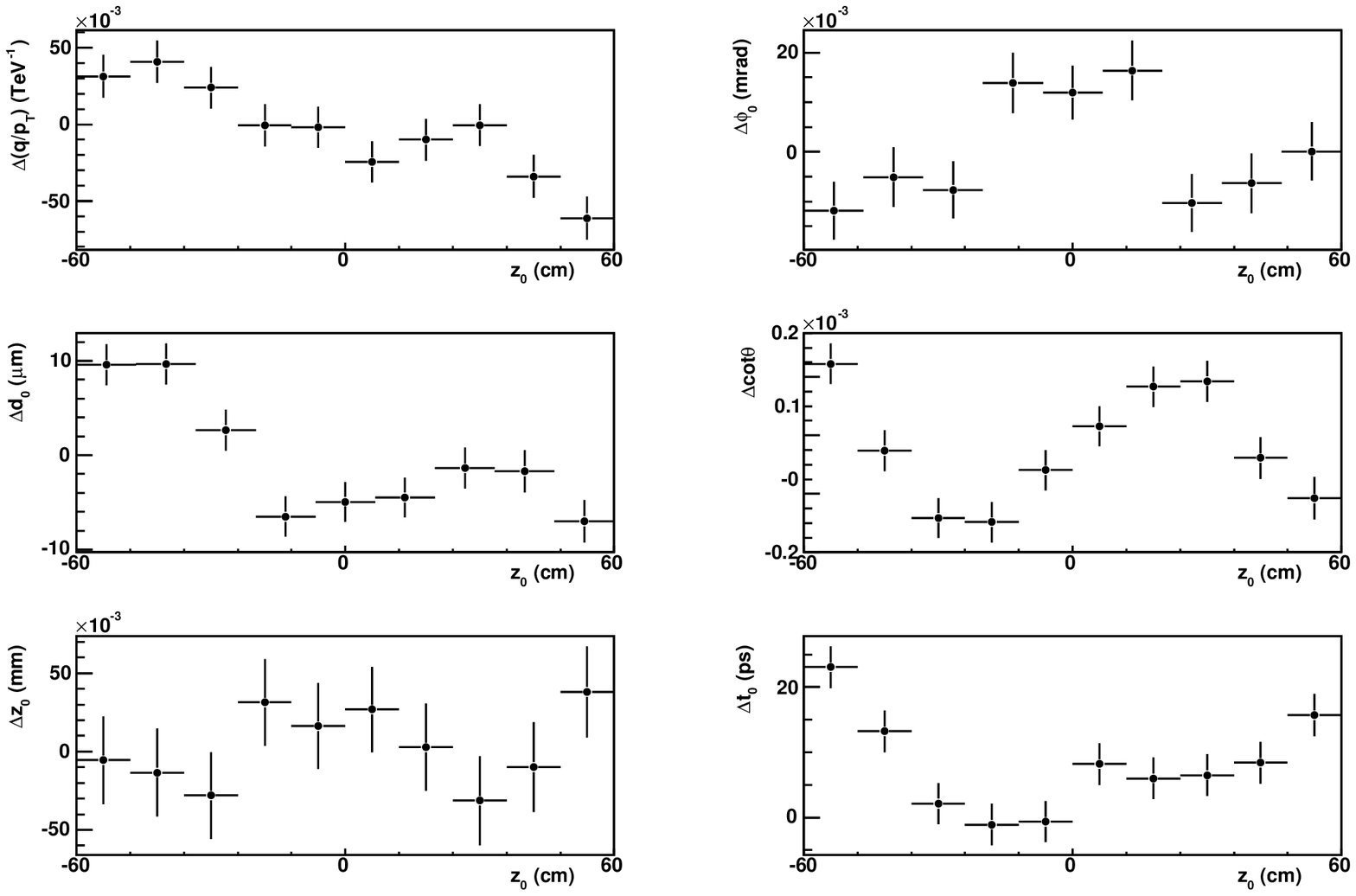}
\end{center}
\caption{Track parameter pulls as functions of $z_0$ after applying the cosmic-ray corrections.   }
\label{fig:pullsVsZ0_after}
\end{figure*}

\begin{figure*}[!tbhp]
\begin{center}
\epsfxsize 15cm
\epsffile {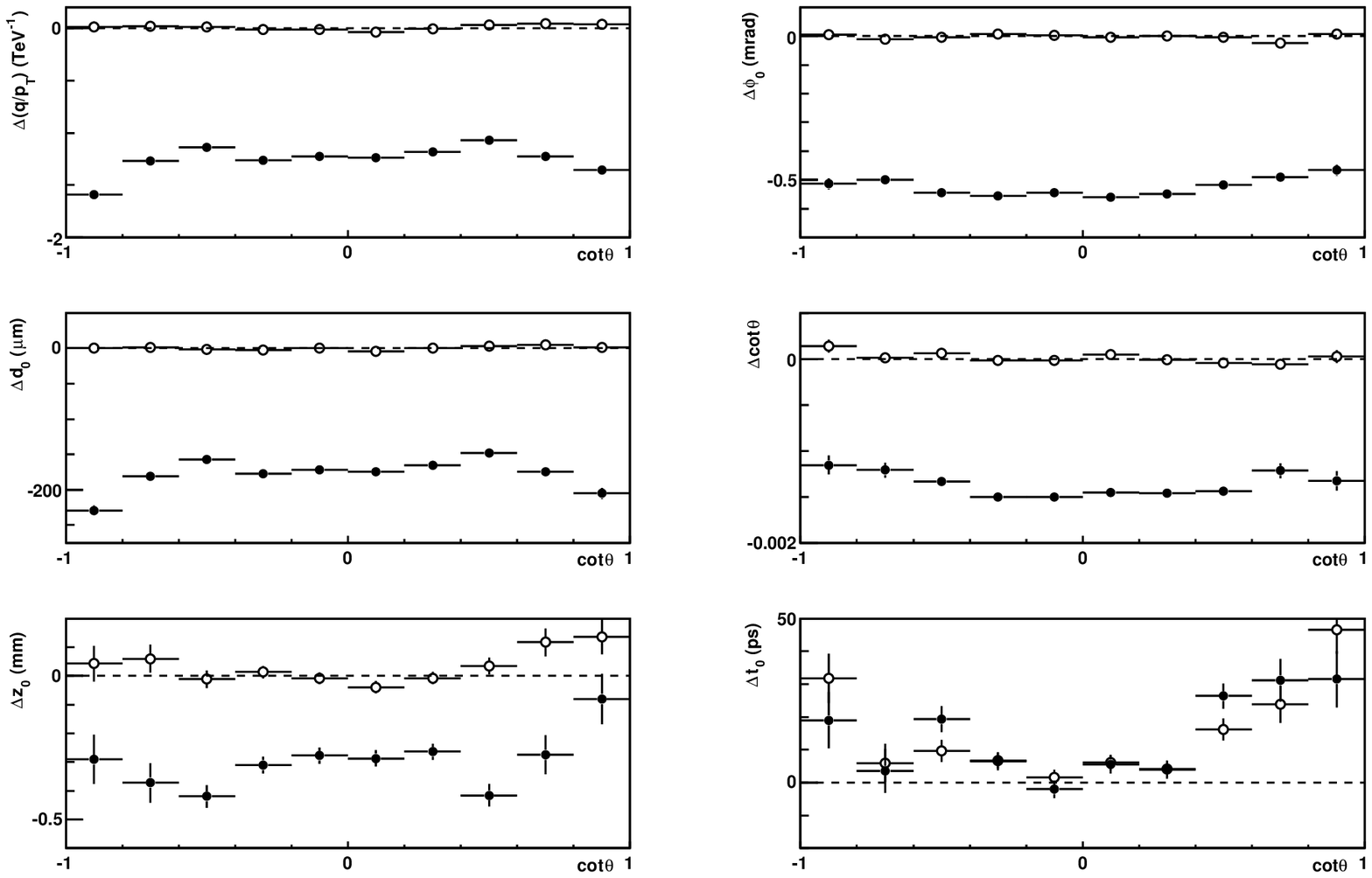}
\end{center}
\caption{Track parameter pulls as functions of $\cot \theta$ before (solid circles) and after (open circles) the cosmic-ray alignment.   }
\label{fig:pullsVsCotThetaOverlay}
\end{figure*}

\begin{figure*}[!tbhp]
\begin{center}
\epsfxsize 15cm
\epsffile {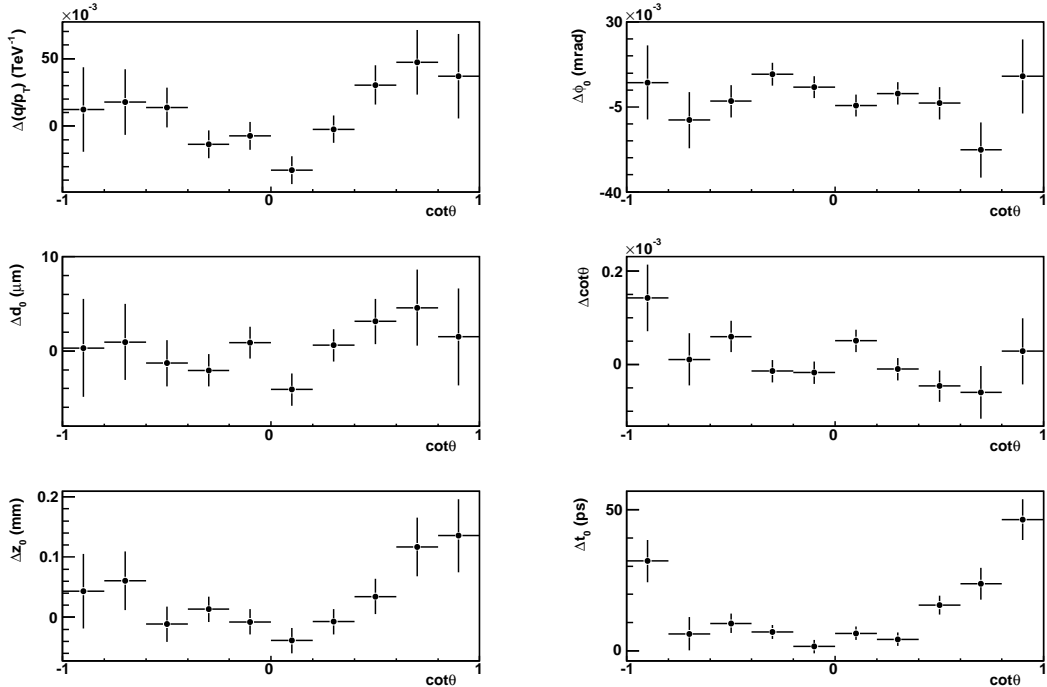}
\end{center}
\caption{Track parameter pulls as functions of $\cot \theta$ after applying the cosmic-ray corrections.   }
\label{fig:pullsVsCotTheta_after}
\end{figure*}

 Overlays of the pulls before and after the cosmic-ray alignment are shown in Fig.~\ref{fig:pullsVsPhi0Overlay}-\ref{fig:pullsVsCotTheta_after}. 

The pulls as a function of azimuth (Fig.~\ref{fig:pullsVsPhi0Overlay}) show substantial variation before alignment.  The variation is largely eliminated by the alignment, as shown again 
 in  Fig.~\ref{fig:pullsVsPhi0_after} on a finer scale. 
The pulls as a function of $z_0$ are also substantially improved by the alignment, as can be seen in Fig.~\ref{fig:pullsVsZ0Overlay}.  The final pulls are shown in Fig.~\ref{fig:pullsVsZ0_after}
 on a finer scale. 
 Similar improvement is seen in the variation of the pulls as a function of polar angle, in the  comparison shown in Fig.~\ref{fig:pullsVsCotThetaOverlay}. A close-up view 
 of these pulls after the alignment is shown in Fig.~\ref{fig:pullsVsCotTheta_after}. 

\section{Conclusions}
The precise alignment of the CDF drift chamber, which provides the basis of the track momentum calibration for the measurement of the $W$ boson mass, is performed using a sample of
 cosmic rays collected {\it in situ} with the collider data. Special reconstruction and track fitting algorithms are employed to associate and fit the hits on both sides of the drift chamber
  with a single helical trajectory. The hit residuals with respect to this reference track provide information on the relative alignment of the drift chamber cells, and on the coherent
 deformations of the drift chamber. Relative rotations of the radial superlayers and relative twists of the endplates are well-constrained by our procedure while, in particular, these modes
 of deformation are not well-constrained by residuals with respect to tracks from $p \bar{p}$ collisions. 

  We align the wires to sub-micron statistical precision, and constrain the differences
 between the endplates to the precision of a few microns. We have also tuned the wire-shape functions describing the deflections due to  electrostatic forces to a similar
 precision. The drift model, and in particular the non-linear time-to-distance relations close to the wire and field sheet, has been updated. 
 
 Remaining biases on  track parameters are, on average, 
  sub-micron on the impact parameter and $<1$\% on the curvature of a $p_T = 1$~TeV 
 particle; the variations with $z_0$, $\phi_0$ and $\cot \theta$ are within a factor of five of the average bias. 

{\bf Acknowledgements} \\
We wish to thank our colleagues on the CDF experiment at the Fermi National Accelerator Laboratory. We thank Aseet Mukherjee, Robert Wagner and Oliver Stelzer-Chilton for useful discussions, 
 and Yu Zeng for his help in isolating the cosmic-ray data. 
We acknowledge the support of the U.S. Department of Energy  
and the Science and Technology Facilities Council of the United Kingdom.


\begin{thebibliography}{99}
\bibitem{conventions}
Pseudorapidity is defined as $\eta = - \ln [\tan (\theta / 2)]$, 
where  $\theta$ is the polar angle from the beam axis. The azimuthal angle is denoted by $\phi$. Energy (momentum) transverse to the beam is denoted as $E_T$ ($p_T$). 
 We use the convention $c = 1$ throughout this paper.
\end{thebibliography}
\end{document}